\documentclass[prb,aps,twocolumn,showpacs]{revtex4-1}
\usepackage{amsmath,amssymb,amsfonts,float}
\usepackage[pdftex]{graphicx,color}    
\usepackage{epstopdf}
\usepackage{times,pifont}
\usepackage{hyperref}
 
\renewcommand{\Vec}[1]{\mathbf{#1}} 

\newcommand{\bolS}{\text{\bf S}}

\newcommand{\bolT}{\text{\bf T}}

\newcommand{\bolOmega}{\mathbf{\Omega}}
\newcommand{\bolA}{\mathbf{A}}
\newcommand{\bolB}{\mathbf{B}}

\newcommand{\bolp}{\mathbf{p}}
\newcommand{\bolq}{\mathbf{q}}
\newcommand{\bolr}{\mathbf{r}}
\newcommand{\bolt}{\mathbf{t}}

\newcommand{\bra}[1]{\langle #1 |}  
\newcommand{\ket}[1]{| #1 \rangle}  
\newcommand{\VEV}[1]{\langle #1 \rangle}  

\newcommand{\be}{\text{e}}

\newsavebox{\dotdot}
\savebox{\dotdot}[3mm]{\shortstack{\circle*{0.8}\\ \\ \circle*{0.8}}}

\begin{document}
\title{%
Semiclassical Approach to Competing Orders in Two-leg Spin Ladder with 
Ring-Exchange
}
\author{K.~Totsuka}
\affiliation{Yukawa Institute for Theoretical Physics, 
Kyoto University, Kitashirakawa Oiwake-Cho, Kyoto 606-8502, Japan}
\author{P.~Lecheminant}
\affiliation{Laboratoire de Physique Th\'eorique et
Mod\'elisation, CNRS UMR 8089,
Universit\'e de Cergy-Pontoise, Site de Saint-Martin, 
2 avenue Adolphe Chauvin, 
95302 Cergy-Pontoise Cedex, France}
\author{S.~Capponi}
\affiliation{Laboratoire de Physique Th{\'e}orique, Universit{\'e} de Toulouse, UPS (IRSAMC), F-31062 Toulouse, France}
\affiliation{CNRS, LPT (IRSAMC), F-31062 Toulouse, France}
\begin{abstract}
We investigate the competition between different orders in the two-leg spin ladder with
a ring-exchange interaction by means of a bosonic approach. 
The latter is defined in terms of spin-1 hardcore bosons which treat the N\'eel and vector chirality
order parameters on an equal footing. A semiclassical approach of the resulting model
describes the phases of the two-leg spin ladder with a ring-exchange.
In particular, we derive the low-energy effective actions which govern
the physical properties of the rung-singlet and dominant vector chirality phases. 
As a by-product of our approach, 
 we reveal the mutual induction phenomenon between spin and chirality with, for instance,
 the emergence of a vector-chirality phase from the application of a magnetic field in bilayer
 systems coupled by four-spin exchange interactions.
\end{abstract}            
\maketitle
\section{Introduction}

Multiple-spin exchange interactions have attracted much interest
for a long time. 
These interactions appears either as many-body direct exchange processes\cite{Thouless-65} 
or as higher-order corrections in the strong-coupling expansion of the half-filled Hubbard model. 
\cite{Takahashi-77,MacDonald-G-Y-88}
In most cases (especially in Mott insulators), the four-spin ring (or cyclic) exchange 
is rather small compared with the usual Heisenberg term. However, it may play a significant role 
in $^{3}$He, where hardcore repulsion makes many-body exchanges more likely than 
the standard two-body one, and Mott insulators close to the metal-insulator transitions. 
In fact, such interactions are expected to be crucial for explaining
unusual magnetic behavior in $^{3}$He absorbed on graphite\cite{Roger-H-D-83,Fukuyama-08}, 
Wigner crystals \cite{Chakravarty-K-N-V-99} and some Mott insulators on triangular lattices.\cite{Shimizu-M-K-M-S-03,*Yamashita-etal-10} 
The relevance of the four-spin cyclic exchange has also been reported%
\cite{Coldea-H-A-P-F-M-C-F-01,Brehmer-M-M-N-U-99,%
Matsuda-K-E-B-M-00,Nunner-B-K-W-G-02,Calzado-G-B-C-M-03,%
Schmidt-U-05,Notbohm-etal-07,Bordas-G-C-C-05,Lake-etal-10,Mizuno-T-M-99} 
in fully describing the inelastic neutron-scattering experiments
for such cuprates as La$_2$CuO$_4$, La$_6$Ca$_8$Cu$_{24}$O$_{41}$,
La$_4$Ca$_{10}$Cu$_{24}$O$_{41}$, CaCu$_2$O$_3$, and SrCu$_2$O$_3$.

A second motivation to investigate multiple-spin exchange interactions stems from 
the exotic physics that emerges from their competition with the Heisenberg
spin exchange. 
For instance, certain sorts of spin nematic phases are known to be stabilized
in  some two-dimensional Heisenberg magnets with ring-exchange interaction.
\cite{Lauchli-D-L-S-T-05,Shannon-M-S-06,Momoi-S-S-06} 
Also multiple-spin exchanges can
lead to new emerging quantum critical behaviors like
the deconfined quantum criticality \cite{Senthil-V-B-S-F-04,*Senthil-B-S-V-F-04,Sandvik-07,*Sandvik-10} or
the spin Bose-metal. \cite{Sheng-M-F-09,*Block-M-K-S-M-F-11} 
On top of these novel phases, even more exotic topological phases have been predicted  
to be realized with such interactions. \cite{Misguich-B-L-W-98,Freedman-N-S-05} 

\begin{figure}[H]
\begin{center}
\includegraphics[scale=0.4]{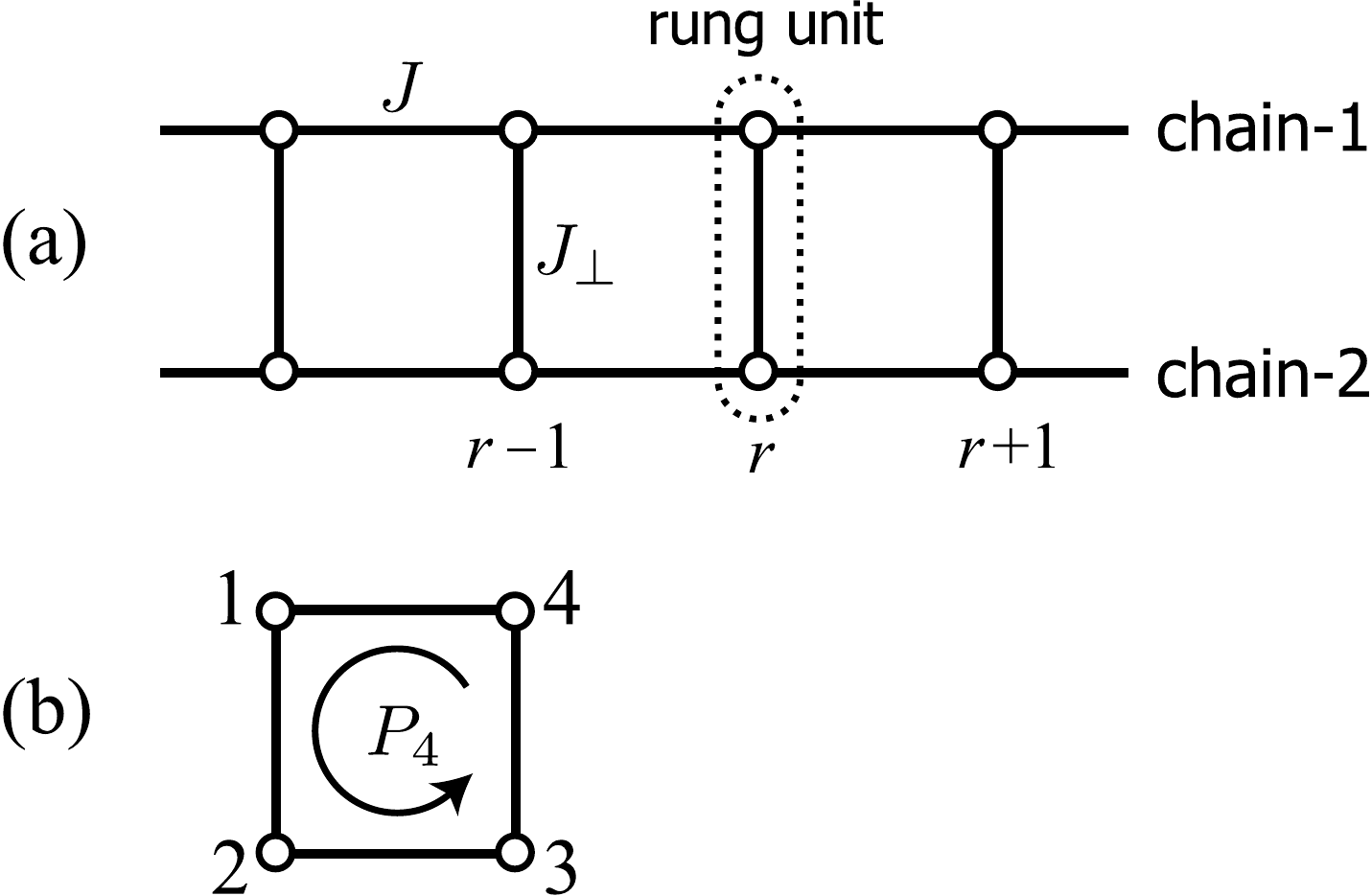}
\caption{Two-leg ladder (a) and four-spin plaquette (b) on which 
ring-exchange $P_{4}$ is defined. %
\label{fig:2leg_ladder}}
\end{center}
\end{figure}

A paradigmatic and minimal model 
that realizes these two aspects of the multiple-spin exchange interactions 
would be the two-leg spin ladder with a ring-exchange:
\begin{equation}
\begin{split}
{\cal H} 
=& J \sum_{r=\text{rungs}} \left(
\bolS_{1,r}{\cdot}\bolS_{1,r+1}+\bolS_{2,r}{\cdot}\bolS_{2,r+1}
\right) \\
& + J_{\perp}\sum_{r}\bolS_{1,r}{\cdot}\bolS_{2,r}
+ K_{4}\sum_{\text{plaquettes}}\left(
P_{4} + P^{-1}_{4}  \right)   \; ,
\end{split}
\label{eqn:ladder}
\end{equation}
where $\bolS_{a,r}$ ($a = 1,2$) denotes the spin-1/2 operator on the chain-$a$ and the rung
$r$ of the spin ladder. The parameters $J$ and $ J_{\perp}$ respectively 
are the intrachain- and the interchain exchange coupling (Fig.~\ref{fig:2leg_ladder} (a)).
The ring exchange $P_4$ is defined on each 
plaquette (two-rung cluster: Fig.~\ref{fig:2leg_ladder} (b)) and 
cyclically permutes the states of the four spins on the plaquette. 
This model is considered as relevant 
to describe the physical properties of ladder compounds e.g. 
La$_4$Ca$_{10}$Cu$_{24}$O$_{41}$ 
(Refs.~\onlinecite{Brehmer-M-M-N-U-99,Matsuda-K-E-B-M-00}) and 
CaCu$_2$O$_3$ (Ref.~\onlinecite{Lake-etal-10}).  
The model (\ref{eqn:ladder}) is interesting in its own right and 
has been studied extensively over the years and 
its phase diagram\cite{Muller-V-M-02,Lauchli-S-T-03,Hikihara-M-H-03,Momoi-H-N-H-03,Gritsev-N-B-04}, ground-state- and 
dynamical properties\cite{Schmidt-M-U-03,Notbohm-etal-07,Hakobyan-08,Nishimoto-A-09}, 
quantum phase transitions\cite{Hijii-Q-N-03,Lecheminant-T-05,Lecheminant-T-06-SU4}, and entanglement properties%
\cite{Song-G-L-06,Maruyama-H-H-09,*Arikawa-T-M-H-09,Li-S-L-Z-11-unpub} 
have been explored by both analytical and numerical approaches.

The zero-temperature phase diagram is rich and six different phases have been 
identified \cite{Lauchli-S-T-03}:
the ferromagnetic phase, the rung-singlet (RS) phase, the staggered dimerized phase, the scalar chirality
phase, the dominant vector chirality (VC) phase, and the dominant collinear spin phase. All phases are gapful
except the ferromagnetic phase.
The four-spin cyclic exchange manifests itself by stabilizing the scalar chirality phase,
which spontaneously breaks the time-reversal symmetry, and the VC phase. 
The latter phase has a unique singlet ground state with a finite gap.
In sharp contrast to the usual RS phase of the two-leg spin ladder,
the lowest triplet excitation of the VC phase is not the standard triplon created by 
the spin operators $\bolS_{a,r}$ but is built from the vector
chirality  $\bolS_{1,r}{\times} \bolS_{2,r}$ as we will show below. 
Consequently, the dominant ground-state correlations of this phase
occur not in the spin-spin channel but rather in the vector-chirality channel.%
\cite{Lauchli-S-T-03,Hikihara-M-H-03}
The relationship between the RS and VC phases can be simply understood by means of a spin-chirality
duality transformation. \cite{Hikihara-M-H-03,Momoi-H-N-H-03}
Under the transformation
\begin{equation}
\begin{split}
& \bolS_{1,r} \rightarrow \frac{1}{2} \left(\bolS_{1,r}  + \bolS_{2,r} \right) - \bolS_{1,r}{\times} \bolS_{2,r}
 \\
& \bolS_{2,r} \rightarrow \frac{1}{2} \left(\bolS_{1,r}  + \bolS_{2,r} \right) + \bolS_{1,r}{\times} \bolS_{2,r}\; ,
\end{split}
\label{spinchirality}
\end{equation}
the N\'eel staggered magnetization $\bolS_{1,r}{-} \bolS_{2,r}$ is interchanged with
the VC order parameter  $2 \bolS_{1,r}{\times} \bolS_{2,r}$ and vice-versa. \cite{Hikihara-M-H-03,Momoi-H-N-H-03}
Model (\ref{eqn:ladder}) is invariant, i.e. self-dual, under the duality transformation (\ref{spinchirality})
when $K_4 = J/2$. 
Physics in the vicinity of this point is dictated by the competition between
the RS and the VC orders.

In this paper, we investigate the competition between these orders by means of a semiclassical
approach which treats the  N\'eel  antiferromagnetic (NAF) 
and the VC order parameters on an equal footing.
To this end, the model (\ref{eqn:ladder}) is first expressed in terms of 
spin-1 hardcore boson operators that create triplet states out of
the vacuum. \cite{Sachdev-B-90}  
In this basis, the spin-chirality transformation (\ref{spinchirality}) has a simple interpretation
as the U(1) gauge transformation of the bosons.\cite{Lecheminant-T-06-SU4}  
The next step of the approach is to perform a simple mean-field approximation 
of the phases of the bosonic model.  
We then incorporate quantum fluctuations by constructing  semiclassical low-energy Hamiltonians 
which describe the competition of the RS and the VC orders as well as 
the usual rotational fluctuations. 
It is important to observe that our approach is not restricted to one dimension but
applies to e.g. higher-dimensional systems consisting of two-spin clusters as well. 
In this respect, the approach is directly relevant to two-dimensional half-filled bilayer spin-1/2 fermions with ring-exchange interaction.\cite{Kolezhuk-07}  
Finally, we investigate the interesting mutual induction phenomenon between spin and chirality
degrees of freedom within our semiclassical approach 
which signals, for instance, the emergence of a VC phase
by the application of a magnetic field. This last result has been first predicted in one dimension
by means of the bosonization approach.\cite{Sato-07}  

The rest of the paper is organized as follows.
In Sec. \ref{sec:effective-spin1-model}, 
the mapping of model (\ref{eqn:ladder}) onto that of spin-1 hardcore bosons 
and the path-integral representation of the resulting model, on which our 
semiclassical approach is based, are presented.  
Then, we draw the semiclassical phase diagram in Sec. \ref{sec:mean-field-PD} 
by minimizing the classical 
energy, which is equivalent to mean-field approximation. 
The low-energy effective actions for the main phases are then derived 
in Sec. \ref{sec:effective-action}.
As an application of our method, we investigate the mutual induction phenomenon 
between spin and chirality
in Sec. \ref{sec:effect-magnetic-field}. 
Finally, our concluding remarks are presented in Sec. \ref{sec:concluding-remark}.
The paper is supplied with one appendix which provides
some technical information on the semiclassical approach.
\section{Effective spin-1 hardcore bosons approach}
\label{sec:effective-spin1-model}
In this section, we introduce three hardcore bosons to map
the model (\ref{eqn:ladder}) onto an effective spin-1 boson Hamiltonian.
This approach will enable us to illustrate how competition among different 
orders (specifically, antiferromagnetic (AF) order and chiral one) 
is described in our framework.  

\subsection{Dimer basis and spin-chirality rotation}
\label{sec:dimer-basis}
First we begin with analyzing the Hilbert space of each rung which is an obvious unit of construction. 
In describing the states on each rung, it is convenient to move from the standard spin-1/2 
($\uparrow$/$\downarrow$) basis to the singlet-triplet basis: 
\begin{equation}
\begin{split}
& |\text{s}\rangle = \frac{1}{\sqrt{2}}(
\ket{\uparrow\downarrow} - \ket{\downarrow\uparrow}) \\
& |\text{t}_{x}\rangle 
= -\frac{1}{\sqrt{2}}(\ket{\uparrow\uparrow}-\ket{\downarrow\downarrow}) 
\; , \; \; 
|\text{t}_{y}\rangle 
=\frac{i}{\sqrt{2}}(\ket{\uparrow\uparrow}+\ket{\downarrow\downarrow}) \\
&|\text{t}_{z}\rangle = \frac{1}{\sqrt{2}}(
\ket{\uparrow\downarrow} + \ket{\downarrow\uparrow}) \; .
\end{split}
\end{equation}
These four states may be thought of as created by the following 
four (constrained) bosons obeying the standard bosonic 
commutation relations:\cite{Sachdev-B-90}
\begin{equation}
\begin{split}
&\ket{\text{s}}=s^{\dagger}\ket{0} \; , \; 
\ket{\text{t}_{\alpha}} = t_{\alpha}^{\dagger}\ket{0} \; \; (\alpha=x,y,z)\\
& s^{\dagger}s + \sum_{\alpha=x,y,z}t^{\dagger}_{\alpha}t_{\alpha} =1 \; .
\end{split}
\end{equation}
For our purpose, however, it is more convenient to identify 
$\ket{s}$ with the empty state (boson vacuum) 
and represent the remaining triplet states by using three {\em hardcore} 
bosons $b_{\alpha}$
\begin{equation}
|0\rangle_{\text{hcb}} \equiv |\text{s}\rangle \; , \; 
b^{\dagger}_{\alpha}|0\rangle_{\text{hcb}} \equiv |\text{t}_{\alpha}\rangle 
\end{equation}
 obeying the following nonstandard commutation relations:
\begin{equation}
 [\, b_{r,\alpha}\, , \, b^{\dagger}_{r^{\prime},\beta}\,] = 
\left\{
\delta_{\alpha \beta}\!\left(1-n^{\text{B}}_{r} \right)
-b_{r,\beta}^{\dagger}b_{r,\alpha}
\right\} \delta_{r,r^{\prime}} \; ,
\end{equation}
with $n_{\text{B}}$ being the number operator of the hardcore boson:
\begin{equation}
n^{\text{B}}_{r} \equiv \sum_{\alpha=x,y,z} b_{r,\alpha}^{\dagger}b_{r,\alpha}
=\bolS_{1,r}{\cdot}\bolS_{2,r}+3/4 \; .
\end{equation}
The crucial step is to realize that under the hardcore constraint 
\begin{equation}
 n^{\text{B}}_{r}  = 0,1 \; ,
\end{equation}
the boson creation operator can be represented by a complex combination of 
two different order parameters:\cite{Lecheminant-T-06-SU4}
\begin{equation}
b^{\dagger}_{r,\alpha} = \frac{1}{2}(\bolS_{1,r}-\bolS_{2,r})^{\alpha}
+ i (\bolS_{1,r}{\times}\bolS_{2,r})^{\alpha} \quad 
(\alpha=x,y,z) \; .
\label{eqn:spin-boson}
\end{equation} 
From this equation, one sees that the competition between 
the AF fluctuations (carried by $(\bolS_{1,r}{-}\bolS_{2,r})$) 
and the (vector) chiral ones ($(\bolS_{1,r}{\times}\bolS_{2,r})$) 
is implemented in a single bosonic object $b$ in a unifying way.  
Also, it is easy to see that (spin-independent) gauge transformation 
$b^{\dagger}_{\alpha}\mapsto \be^{i\varphi}b^{\dagger}_{\alpha}$ translates 
into an SO(2) rotation ({\em spin-chirality rotation}\cite{Momoi-H-N-H-03}) 
for the order-parameter doublet 
$(\bolS_{1,r}-\bolS_{2,r},\bolS_{1,r}{\times}\bolS_{2,r})$:
\begin{equation}
\begin{pmatrix}
\Vec{S}_{1}{-}\Vec{S}_{2} \\
2\Vec{S}_{1}{\times}\Vec{S}_{2}
\end{pmatrix}
\mapsto 
\begin{pmatrix}
\cos\varphi & -\sin\varphi \\
\sin\varphi & \cos\varphi 
\end{pmatrix}
\begin{pmatrix}
\Vec{S}_{1}{-}\Vec{S}_{2} \\
2\Vec{S}_{1}{\times}\Vec{S}_{2}
\end{pmatrix} \; .
\label{eqn:O2doublet}
\end{equation} 
In particular, the spin-chirality duality transformation (\ref{spinchirality}) can be 
obtained from Eq. (\ref{eqn:O2doublet}) with $\varphi = \pi/2$.

\subsection{Generalized boson Hubbard model}
\label{sec:spin1-BH}
It is convenient to rewrite the model Hamiltonian 
(\ref{eqn:ladder}) in terms of 
the hardcore boson introduced above. 
In terms of the bosonic operators $b_{r,\alpha}$, 
the (spin) Hamiltonian (\ref{eqn:ladder}) 
can be mapped onto the following spin-1 Bose-Hubbard model:\cite{Lecheminant-T-06-SU4}
\begin{equation}
\begin{split}
{\cal H} 
= &  \; t \sum_{r,\alpha}
\left(b_{r,\alpha}^{\dagger}b_{r+1,\alpha}+b_{r+1,\alpha}^{\dagger}b_{r,\alpha}\right) \\
& + u \sum_{r,\alpha}
\left(b_{r,\alpha}^{\dagger}b^{\dagger}_{r+1,\alpha}+b_{r+1,\alpha}b_{r,\alpha}\right)  \\
& +  \sum_{r}\left[
J_{\text{bl}}\bolT_{r}{\cdot}\bolT_{r+1}
+ J_{\text{bq}} \left(\bolT_{r}{\cdot}\bolT_{r+1}\right)^{2}\right] \\ 
& + V_{\text{c}} \sum_{r}n_{r}^{\text{B}}n_{r+1}^{\text{B}}
- \mu \sum_{r}n_{r}^{\text{B}}
 \; .
\end{split}
\label{eqn:spin1tJ}
\end{equation}
The spin ${\bf T}$ of the hardcore boson $b$ is given by
\begin{equation}
{\bf T}_{r,\alpha} \equiv -i\, \epsilon_{\alpha \beta \gamma}b^{\dagger}_{r,\beta}b_{r,\gamma} 
\label{spin1op}
\end{equation}
where the projection onto the occupied (i.e. $n^{\text{B}}\neq 0$) states 
is implied on both sides.  
The ring-exchange model (\ref{eqn:ladder}) is reproduced if we choose:
\begin{equation}
\begin{split}
& t=\frac{1}{2}J+K_{4}\; , \;\;
u=\frac{1}{2}J-K_{4}\; , \;\; 
J_{\text{bl}}=\frac{1}{2}J+K_{4}\; , \\
& J_{\text{bq}}=0 \; , \;\;
V_{\text{c}}=4K_{4}\; , \;\; \mu=4K_{4}-  J_{\perp} \; .
\end{split}
\label{eqn:laddercoupling}
\end{equation}
Note that the Hamiltonian (\ref{eqn:spin1tJ}) is 
the most general one allowed by the requirements of 
(i) SU(2) symmetry, (ii) time-reversal invariance, (iii)  exchange of the two chains 
($1\leftrightarrow 2$) and (iv) short-range interactions (i.e. interactions only involve two 
adjacent rungs).  In this respect, model (\ref{eqn:spin1tJ}) describes two-leg spin ladder
with general four-spin exchange interactions.\cite{Lecheminant-T-06-SU4}
The ring-exchange is special in the sense that there is no biquadratic exchange interaction:
$J_{\text{bq}} = 0$.
It is also important to note that the U(1) gauge-symmetry for the bosons, i.e. the spin-chirality 
rotation (\ref{eqn:O2doublet}), is {\em explicitly} 
broken unless $u=0$. The spin-chirality duality symmetry (\ref{spinchirality}) transforms 
the pairing term as: $u \mapsto -u$.  
In the self-dual case, i.e. $u=0$,  model  (\ref{eqn:spin1tJ})  is directly relevant to
spinor Bose quantum gases with hyperfine spin $F=1$ loaded into an optical lattice. \cite{Ho-98,*Ohmi-M-98,*Imambekov-L-D-03} 

\label{sec:Dimer-coherent-state}
In the following sections, we develop a semiclassical approach 
to discuss the physical properties of various phases 
of the model (\ref{eqn:ladder}) (or the equivalent model 
(\ref{eqn:spin1tJ})).  
In particular, we are interested in 
the two non-magnetic phases\cite{Lauchli-S-T-03,Hikihara-M-H-03} 
(dubbed `rung singlet' and `dominant vector chirality' in Ref.~\onlinecite{Lauchli-S-T-03}). 
The effective models derived in Sec. \ref{sec:effective-action} provide us with a simple
and natural framework of describing the competition of spin and 
chirality degrees of freedom.  
\subsection{Coherent-state-construction of path-integral}
As usual, the starting point of the path-integral approach is the construction of the many-body coherent state basis
for the two-leg ladder model with a ring-exchange (\ref{eqn:ladder}) and the spin-1 boson Hubbard 
model (\ref{eqn:spin1tJ}). 
In this respect, let us consider spin-1 hardcore bosons $b^{\dagger}_{\alpha}$ 
$(\alpha=x,y,z)$ on  a unit of construction (e.g. a rung in the case of the two-leg ladder).  
As has been mentioned in Sec. II A, 
these boson operators are parametrized in terms of 
the two competing (real) order parameters   
$\hat{\bf q}$ and $\hat{\bf p}$:
\begin{equation}
b^{\dagger}_{r,\alpha} = (\hat{\bf q}_{r})_{\alpha} + i (\hat{\bf p}_{r})_{\alpha} 
\;\; (\alpha=x,y,z) \; ,
\label{eqn:boson-by-q-p}
\end{equation}
where in our ladder problem, $(\hat{\bf q}_{r})_{\alpha}$ and $(\hat{\bf p}_{r})_{\alpha}$
are respectively the  staggered magnetization and
the vector-chiral order parameter:
\begin{equation}
\hat{\mathbf{q}}_{r}=\frac{1}{2}(\bolS_{1,r} {-} \bolS_{2,r}) \; , \;\;
\hat{\mathbf{p}}_{r}= \bolS_{1,r}{\times}\bolS_{2,r} 
\end{equation}
(see Eq. (\ref{eqn:spin-boson})).

Clearly, arbitrary states on each unit can be represented 
as:\cite{Kolezhuk-96} 
\begin{equation}
|\psi\rangle = s |\text{s}\rangle + \sum_{\alpha=x,y,z}
t_{\alpha}|\text{t}_{\alpha}\rangle 
\equiv | s, \boldsymbol{t} \rangle\; ,  
\end{equation}
where  
\begin{equation}
s^{\ast}s+ \boldsymbol{t}^{\ast}{\cdot}\boldsymbol{t}=1
\; ,
\label{eqn:normalization}
\end{equation}
is required by the normalization condition. 
Since the overall phase is irrelevant, we may fix the gauge 
in such a way that the singlet amplitude $s$ is real and positive: 
\begin{equation}
s = \sqrt{1-\boldsymbol{t}^{\ast}{\cdot}\boldsymbol{t}} 
\quad 
(\boldsymbol{t}^{\ast}{\cdot}\boldsymbol{t}\leq 1) 
\end{equation}
and parametrize the complex vector $\boldsymbol{t}$ in terms of 
two {\em real} vectors $\bolA$ and $\bolB$ as:
\begin{equation}
\begin{split}
& \boldsymbol{t} \equiv \bolA - i \, \bolB  \quad , \quad 
s = \sqrt{1-(\bolA{\cdot}\bolA+\bolB{\cdot}\bolB)}  \\
& (\bolA,\bolB\in \mathbb{R}^{3}\, , \,
\bolA{\cdot}\bolA + \bolB{\cdot}\bolB\leq 1)
\; .
\end{split}
\label{eqn:gauge-fixed}
\end{equation}

One of the greatest merits of using this representation is that the two 
order parameters 
$\hat{\bolq}$ and $\hat{\bolp}$, which are related to the boson by 
Eq. (\ref{eqn:boson-by-q-p}),
are expressed simply by $\bolA$ and $\bolB$: 
\begin{subequations}
\begin{align}
& \langle s,\boldsymbol{t}|\hat{\bf q}|s,\boldsymbol{t}\rangle
= (s^{\ast}\boldsymbol{t}+\boldsymbol{t}^{\ast}s)/2 
= \sqrt{1-(\bolA^{2}+\bolB^{2})}\, \bolA \\
& \langle s,\boldsymbol{t}|\hat{\bf p}
|s,\boldsymbol{t}\rangle
= i(s^{\ast}\boldsymbol{t} - \boldsymbol{t}^{\ast}s)/2  
= \sqrt{1-(\bolA^{2}+\bolB^{2})}\, \bolB
\; .
\end{align}
\end{subequations}
From these equations, it is obvious that the spin-chirality rotation  
(\ref{eqn:O2doublet}) is equivalent to the {\em gauge transformation} 
of the triplet operators $\boldsymbol{t}$ 
\begin{equation} 
\boldsymbol{t}^{\ast} \mapsto \be^{i\varphi}\boldsymbol{t}^{\ast}
\label{eqn:gauge-tr}
\end{equation}
or the following O(2) transformation for the pair $(\bolA,\bolB)$:
\begin{equation}
\begin{pmatrix} 
\bolA \\ \bolB
\end{pmatrix} \mapsto 
\begin{pmatrix}
\cos \varphi & -\sin \varphi \\
\sin\varphi & \cos \varphi 
\end{pmatrix}
\begin{pmatrix} 
\bolA \\ \bolB
\end{pmatrix}  \; .
\label{eqn:spin-chirality-by-AB}
\end{equation} 
In the case of the two-leg ladder where the triplet boson is defined on 
the rung, 
the pair of order parameters is given by Eq.~(\ref{eqn:spin-boson}) 
and the above is nothing but the spin-chirality 
transformation (\ref{eqn:O2doublet}).\cite{Hikihara-M-H-03}

The magnetic moment ${\bf T}$, that generates the O(3) rotation 
of the boson triplet, is expressed as 
(see Eq.~(\ref{spin1op})):
\begin{equation}
\langle s,\boldsymbol{t}| \hat{\bf T} |s,\boldsymbol{t}\rangle 
= -i(\boldsymbol{t}^{\ast}{\times}\boldsymbol{t}) 
= - 2(\bolA{\times}\bolB) \; .
\label{eqn:local-moment-by-AB}
\end{equation}
If we take a single rung as the unit in the case of the two-leg ladder, 
$\hat{\bf T}=\bolS_{1}{+}\bolS_{2}$.  
The boson density is another important quantity and takes 
the following expression:
\begin{equation}
\langle s, \boldsymbol{t} | n^{\text{B}} |s,\boldsymbol{t} \rangle 
= \boldsymbol{t}^{\ast}{\cdot}\boldsymbol{t} 
= (\bolA^{2}+\bolB^{2}) \; . 
\end{equation}
The many-body coherent state basis are constructed as the tensor-product 
of the local coherent states:
\begin{equation}
\bigotimes_{r} |s_{r}, \boldsymbol{t}_{r}\rangle 
= \bigotimes_{r} |\bolA_{r},\bolB_{r}\rangle \; .
\end{equation}

With these expressions, we can write down the desired path-integral 
formula for many-spin systems\cite{Kolezhuk-96} by 
following the standard steps:\cite{Auerbach-book,Wen-book-04}
\begin{equation}
{\cal S} = 
\int\!dt \sum_{r}
(\bolA_{r}{\cdot}\partial_{t}\bolB_{r}
-\bolB_{r}{\cdot}\partial_{t}\bolA_{r}) 
- \int\!dt\, {\cal H}_{\text{cl}}(\{\bolA\},\{\bolB\}) \; ,
\label{action}
\end{equation}
where ${\cal H}_{\text{cl}}$ is given by the expectation value 
of the Hamiltonian with respect to the many-body coherent state
$\otimes_{r}|\text{s}_{r},{\bf t}_{r}\rangle$. 
In obtaining ${\cal H}_{\text{cl}}$, 
the easiest way is to use the spin-1 Bose-Hubbard model (\ref{eqn:spin1tJ}) 
with the coupling constants (\ref{eqn:laddercoupling}) and 
then plug the following expressions:
\begin{equation}
\begin{split}
& b^{\dagger}_{r,\alpha} \mapsto 
\sqrt{1-(\bolA_{r}^{2}+\bolB_{r}^{2})}\,
(\bolA_{r} +i\, \bolB_{r}) \; , \\
& b_{r,\alpha} \mapsto \sqrt{1-(\bolA_{\bolr}^{2}+\bolB_{r}^{2})}\,
(\bolA_{r} -i\, \bolB_{r})  \; .
\end{split}
\label{eqn:boson-by-AB}
\end{equation} 
The final result reads as follows:
\begin{subequations}
\begin{align}
&{\cal H}_{\text{cl}} =
{\cal H}_{\text{magnetic}}+{\cal H}_{\text{hopping}}
+{\cal H}_{\text{charge}} \label{eqn:classical-AB} \\
\begin{split}
&{\cal H}_{\text{magnetic}} \\
& \equiv (2J+4K_{4})\sum_{r} \bigl\{
(\bolA_{r}{\cdot}\bolA_{r+1})(\bolB_{r}{\cdot}\bolB_{r+1}) \\
& \phantom{\equiv (2J+4K_{4})\sum_{r} \bigl\{ }\qquad 
-(\bolA_{r}{\cdot}\bolB_{r+1})(\bolB_{r}{\cdot}\bolA_{r+1})
\bigr\} ,
\end{split}
\label{eqn:classical-AB-magnetic}
\\
\begin{split}
& {\cal H}_{\text{hopping}} \\
& \equiv 
\sum_{r} 4\sqrt{1-(\bolA_{r}^{2}+\bolB_{r}^{2})}
\sqrt{1-(\bolA_{r+1}^{2}+\bolB_{r+1}^{2})} \\
& \qquad \qquad \qquad \times
\left\{ \frac{J}{2} \bolA_{r}{\cdot}\bolA_{r+1} 
+ K_{4}\, \bolB_{r}{\cdot}\bolB_{r+1} \right\} ,
\end{split}
\\
\begin{split}
& {\cal H}_{\text{charge}} \equiv 
 (J_{\perp}+2K_{4})\sum_{r}\left\{
(\bolA_{r}^{2}+\bolB_{r}^{2})-\frac{3}{4}
\right\} \\
& 
+ 4K_{4}\sum_{r}
\left\{(\bolA_{r}^{2}+\bolB_{r}^{2})-\frac{3}{4}\right\}
\left\{(\bolA_{r+1}^{2}+\bolB_{r+1}^{2})-\frac{3}{4}\right\} \; .
\end{split}
\label{eqn:classical-AB-charge}
\end{align}
\end{subequations}
The magnetic- and the charge part are invariant under the spin-chirality U(1) 
transformation (\ref{eqn:O2doublet}) and the competition between 
the $(\pi,\pi)$ antiferromagnetic correlation ($\bolA$) 
and the chirality correlation ($\bolB$) is controlled solely 
by the two coupling constants in $\mathcal{H}_{\text{hopping}}$:
\begin{equation}
J_{\text{A}}\equiv \frac{t+u}{2}=\frac{J}{2} \; , \;\; 
J_{\text{B}}\equiv \frac{t-u}{2}=K_{4} \; .
\label{eqn:Ja-Jb}
\end{equation}
\section{Mean-Field Phase Diagram}
\label{sec:mean-field-PD}
In this section, we determine the classical ground state of the model (\ref{eqn:ladder}) 
or its bosonic equivalent (\ref{eqn:spin1tJ}), on the basis of which we develop 
the effective field theories in Sec. \ref{sec:effective-action}. 
\subsection{General properties}
Before presenting the results, we describe the relationship between 
the obtained $\{\bolA,\bolB\}$-configurations and the physical phases. 
By construction, it is obvious that our calculation is nothing but the mean-field 
approximation for the spin-1 boson using the following product state:
\begin{equation}
\begin{split}
& \ket{\Psi}_{\{\eta_{r},\mathbf{a}_{r},\mathbf{b}_{r}\}} \\
&= \bigotimes_{r}\left\{
\cos\frac{\eta_{r}}{2}\ket{\text{s}} 
+ \sin\frac{\eta_{r}}{2}\sum_{\alpha=x,y,z}(\mathbf{a} -i\,\mathbf{b})_{r,\alpha}\ket{\text{t}_{r,\alpha}}
\right\}  \; ,
\end{split}
\label{eqn:MF-wf}
\end{equation}
where we have introduced 
\begin{equation}
s_{r}=\cos\frac{\eta_{r}}{2} \, , \; 
\boldsymbol{t}_{r}^{\ast} 
=\bolA_{r}+i\,\bolB_{r} 
= \sin\frac{\eta_{r}}{2} (\mathbf{a}_{r} + i\, \mathbf{b}_{r}) ,
\end{equation}
with $\mathbf{a}_{r}^{2}+\mathbf{b}_{r}^{2} =1$. 
The parameter $\eta_{r}$ ($0\leq \eta_{r}\leq \pi$) 
controls the local boson density $n_{\text{B}}$ through the relation 
\begin{equation}
\bra{\Psi}
n^{\text{B}}_r \ket{\Psi}_{\{\eta_{r},\mathbf{a}_{r},\mathbf{b}_{r}\}}
= \sin^{2}\frac{\eta_{r}}{2} \; .
\end{equation}

As has been mentioned in the previous section, 
the spin-1 (hardcore) boson operators are parametrized in terms of 
the two competing (real) order parameters   
$\hat{\bf q}$ and $\hat{\bf p}$ (see Eq. (\ref{eqn:boson-by-q-p})).
One can then compute the expectation values of these operators
in the coherent state (\ref{eqn:MF-wf}):
\begin{equation}
\begin{split}
&\bra{\Psi}\hat{q}_{r,\alpha}
\ket{\Psi}_{\{\eta_{r},\mathbf{a}_{r},\mathbf{b}_{r}\}}
= \Bigl \langle \frac{1}{2}(\bolS_{1}{-}\bolS_{2})^{\alpha}\Bigr\rangle 
= \frac{1}{2}\sin\eta_{r}\, (\mathbf{a}_{r})_{\alpha} \; , \\
& \bra{\Psi}\hat{p}_{r,\alpha}
\ket{\Psi}_{\{\eta_{r},\mathbf{a}_{r},\mathbf{b}_{r}\}}
= \bigl\langle (\bolS_{1}{\times}\bolS_{2})^{\alpha}\bigr\rangle
= \frac{1}{2}\sin\eta_{r}\, (\mathbf{b}_{r})_{\alpha}  \; .
\end{split}
\label{eqn:EV-n-p}
\end{equation}
Therefore, the phases with $\eta_{r}\neq 0, \pi$ in general correspond to superfluids 
in that $\VEV{b^{\dagger}_{r,a}}\neq 0$. 

Information on magnetism may be obtained by the spin-1 magnetic moment  
(\ref{spin1op}) on each rung:  
\begin{equation}
\bra{\Psi}{\bf T}_{r}
\ket{\Psi}_{\{\eta_{r},\mathbf{a}_{r},\mathbf{b}_{r}\}}
=\bigl\langle \bolS_{1}{+}\bolS_{2}\bigr\rangle
= -2\sin^{2}\frac{\eta_{r}}{2}
\mathbf{a}_{r}{\times}\mathbf{b}_{r}  \; .
\label{eqn:T-by-beta}
\end{equation}
From Eqs. (\ref{eqn:EV-n-p}) and (\ref{eqn:T-by-beta}), it is obvious that 
if, for some reasons, the system chooses the state with $\mathbf{a}=\mathbf{0}$, 
$\mathbf{b}\neq\mathbf{0}$  
($\eta\neq 0,\pi$), 
a moment free (i.e. $\VEV{\bolS_{1}{+}\bolS_{2}}=\VEV{\bolS_{1}{-}\bolS_{2}}={\bf 0}$) 
chiral phase 
$\VEV{\bolS_{1,r}{\times}\bolS_{2,r}}\neq \mathbf{0}$ 
($p$-type spin-nematic\cite{Andreev-G-85}) 
is realized. 
Similarly, $\mathbf{a}\neq \mathbf{0}$, $\mathbf{b}=\mathbf{0}$ implies 
another type of phases with collinear spin order (either $(\pi,\pi)$ or $(0,\pi)$).  
 
Since we are dealing with spin-1 bosons, we may expect 
(typically for large enough  biquadratic interaction $J_{\text{bq}}$) 
a phase characterized by the following rank-2 tensor ({\em spin-nematic phase}) to occur:
\begin{equation}
Q^{\alpha \beta}_{r} \equiv \frac{1}{2}(T^{\alpha}_{r}T^{\beta}_{r}+T^{\beta}_{r}T^{\alpha}_{r}) 
- \frac{1}{3}\delta^{\alpha \beta}\bolT_{r}^{2} \; .
\end{equation} 
In the mean-field state (\ref{eqn:MF-wf}), the above tensor order parameter takes 
the value:
\begin{equation}
\begin{split}
& \bra{\Psi} Q^{\alpha \beta}_{r}
\ket{\Psi}_{\{\eta_{r},\mathbf{a}_{r},\mathbf{b}_{r}\}} \\
&= -\sin^{2}\frac{\eta_{r}}{2}\left\{
(\mathbf{a}_{r})_{\alpha}(\mathbf{a}_{r})_{\beta}+(\mathbf{b}_{r})_{\alpha}(\mathbf{b}_{r})_{\beta} 
- \frac{1}{3}\delta^{\alpha \beta}
\right\} \; .
\end{split}
\end{equation}
When $\mathbf{a}$ and $\mathbf{b}$ are parallel to each other, 
the spin sector is in general spin-nematic with 
vanishing magnetic moment $\VEV{\mathbf{T}}=\mathbf{0}$.   
Note that the two $\VEV{\mathbf{T}}=\mathbf{0}$ phases 
described above have finite $\VEV{Q^{\alpha \beta}}$ 
and that, in a sense, they may be thought of as spin-nematic.  
The only difference between spin-1 nematic and the other two states 
comes from the filling-dependent overall factors $\sin\eta$ and 
$\sin^{2}(\eta/2)$; the former can exist even in the spin-1 limit $\eta=\pi$ while 
the latter are not.   

The charge part (Eq.~(\ref{eqn:classical-AB-charge})) dictates the charge density 
distribution. For instance, for sufficiently large (positive) $\mu$, 
the density saturates $n_{\text{B}}=1$ and the system reduces to the (localized) 
spin-1 chain.  For large enough $V_{\text{c}}$, on the other hand, 
the system may develop 
inhomogeneity, i.e. form a charge-density wave with alternating  
$n_{\text{B}}=0$ and $n_{\text{B}}=1$. 

The general mean-field phase diagram, which results from these equations, 
will be presented elsewhere.\cite{Totsuka-C-L-unpub-12}  Here, we only stress that 
the set of coupling constants $(t,u)$ is crucial for the competition between 
antiferromagnetism and chirality.  
To see this, we first calculate the kinetic energy by using the mean-field ansatz 
(\ref{eqn:MF-wf}):
\begin{equation}
\begin{split}
& \Bigl\langle 
t \sum_{r,\alpha}
\left(b_{r,\alpha}^{\dagger}b_{r+1,\alpha}+b_{r+1,\alpha}^{\dagger}b_{r,\alpha}\right)
\Bigr\rangle \\
& + \Bigl\langle 
u \sum_{r,\alpha}
\left(b_{r,\alpha}^{\dagger}b^{\dagger}_{r+1,\alpha}+b_{r+1,\alpha}b_{r,\alpha}\right) 
\Bigr\rangle \\
&= \sum_{r}
 \sin\eta_{r}
\sin\eta_{r+1}
\left\{
J_{\text{A}} \,
\mathbf{a}_{\boldsymbol{r}}{\cdot}\mathbf{a}_{r+1}
+ J_{\text{B}} \, 
\mathbf{b}_{r}{\cdot} \mathbf{b}_{r+1}
\right\}  \; ,
\end{split}
\end{equation}
where the two couplings, that  characterize the anisotropy in the spin-chirality space,
are given by Eq. (\ref{eqn:Ja-Jb}).
For $u\neq 0$ and $\eta\neq 0,\pi$, an anisotropic superfluid forms; 
$\mathbf{a}_{r}(\propto \text{Re}(\boldsymbol{t}^{\ast}_{r}))$ 
is dominant when $|J_{\text{A}}|>|J_{\text{B}}|$, while  
$\mathbf{b}_{r}(\propto \text{Im}(\boldsymbol{t}^{\ast}_{r}))$ 
is dominant when $|J_{\text{A}}|<|J_{\text{B}}|$.  

So far, we have presented the mean-field description of the phases of 
the spin-1 boson model.  However,  in one dimension, strong quantum fluctuations 
may destroy the ordered states predicted by the mean-field theory.  
In fact, as we will show in the next section by using low-energy effective theories, 
some of the ordered phases 
are replaced by gapped short-range phases which do not break rotational symmetry.  
\subsection{Two-leg ladder}
\begin{figure}[h]
\begin{center}
\includegraphics[scale=0.45]{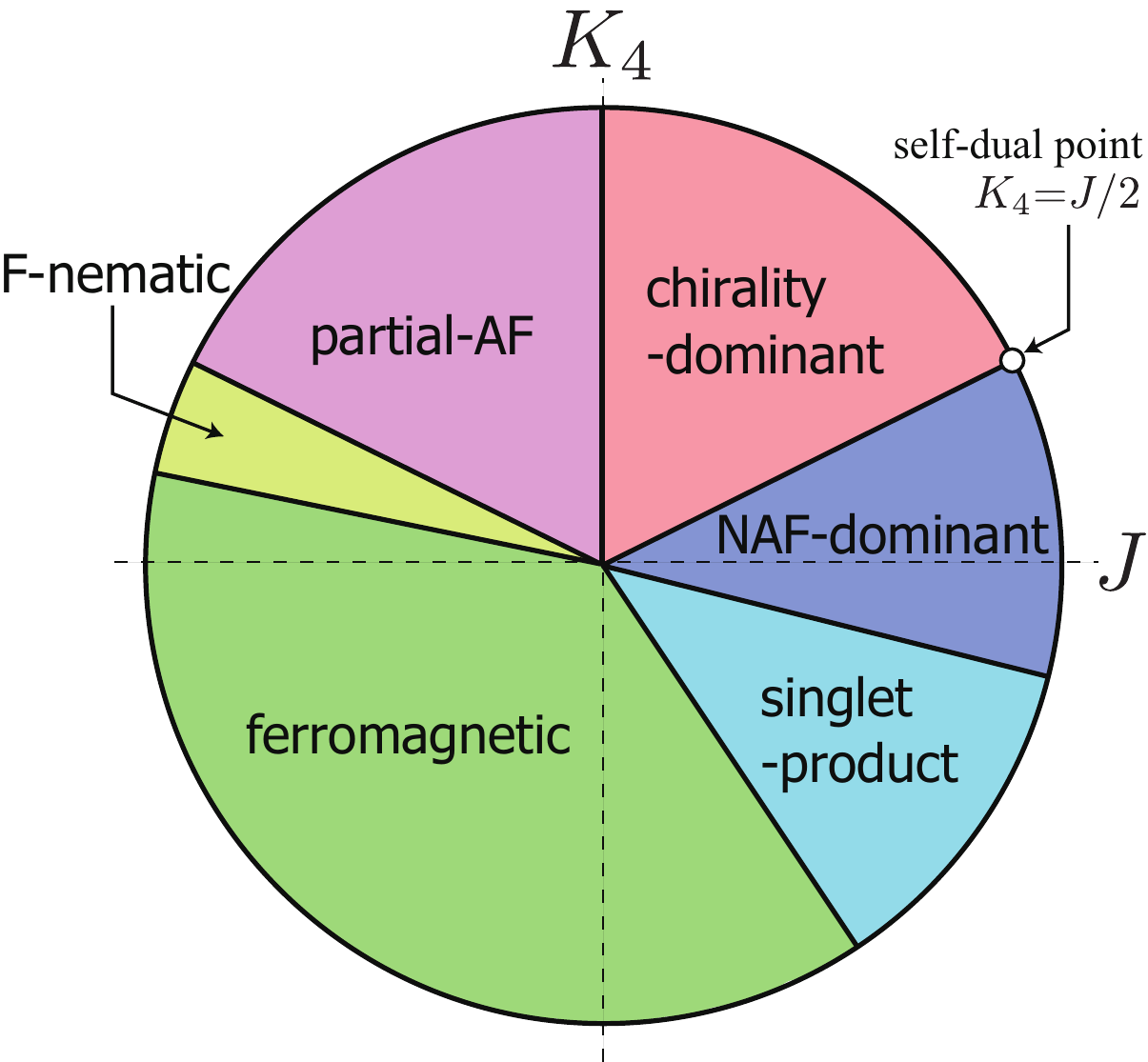}
\caption{(color online) 
Phase diagram of the model (\ref{eqn:ladder}) 
($J=\cos\theta$, $K_{4}=\sin\theta$) obtained 
by applying a mean-field approximation to ${\cal H}_{\text{cl}}$ 
(\ref{eqn:classical-AB}).  
The self-dual point ($K_{4}=J/2$) separates the `NAF-dominant' 
(i.e. $\bolA$-ordered) phase and the `chirality-dominant'  
($\bolB$-ordered) one.  
In $(1{+}1)$-D, strong quantum fluctuations turn these ordered phases into short-ranged 
ones. 
\label{fig:classical-AB-phases}}
\end{center}
\end{figure}

The semiclassical ground state of  the model (\ref{eqn:ladder}) 
(or equivalently, (\ref{eqn:spin1tJ})) is  
obtained by minimizing the energy functional 
$\mathcal{H}_{\text{cl}}(\{\bolA\},\{\bolB\})$ in (\ref{eqn:classical-AB}) 
with respect to the variational parameters $\{\bolA_{r}\}$ and $\{\bolB_{r}\}$.   
As we have already seen, this is nothing but the mean-field treatment 
using (\ref{eqn:MF-wf}).  
The resulting phase diagram contains six phases: (i) `NAF-dominant', 
(ii) `chirality-dominant', (iii) `partial-AF', (iv) `F-nematic', (v) `ferromagnetic' and 
(vi) `singlet-product'. 
It is convenient to parametrize the coupling constants 
as: 
\begin{equation}
J=\cos\theta \; , \;\; K_{4}=\sin\theta \;\; (-\pi < \theta \leq \pi)
\end{equation}
and map out the phase diagram 
as a function of $\theta$ (see Fig.~\ref{fig:classical-AB-phases}).  

Let us describe the nature of the six phases.  \\
(i) {\em NAF-dominant}: 
This phase is described by $\bolA_{r} \ne \mathbf{0}$ and $\bolB_{r} = \mathbf{0}$.
The resulting phase is characterized by vanishing (total) magnetic moment 
on each rung $\VEV{\bolS_{1}{+}\bolS_{2}}=-2\bolA{\times}\bolB$ 
and {\em anti-parallel} ordering of $\bolA$ (local chirality $\bolB=\mathbf{0}$). 
At the mean-field level, the symmetry $\mathrm{SU(2)}{\times}\mathbb{Z}_{2}$ of the model 
(\ref{eqn:ladder}) is broken down to SO(2) (rotation around the ordered $\mathbf{A}$).  
Since $\bolA$ corresponds to $\bolS_{1}{-}\bolS_{2}$, the staggered 
order of $\bolA$ implies the standard $(\pi,\pi)$ N\'{e}el-ordered phase. \\
(ii) {\em chirality dominant}: 
The second phase is the dual ($\mathbf{A}\leftrightarrow \mathbf{B}$)  
of the former with $\bolA_{r} = \mathbf{0}$ and $\bolB_{r} \ne \mathbf{0}$.
This phase has the same symmetry as the first phase and is non-magnetic 
(i.e. $\VEV{\bolS_{1}}=\VEV{\bolS_{2}}=\mathbf{0}$). 
However, the long-range order occurs in the chirality channel 
$\bolB \sim \bolS_{1}{\times}\bolS_{2}$ in a staggered manner.  
The transition from the first phase occurs at the so-called self-dual point 
$\theta =\theta_{\text{sd}}\equiv \tan^{-1}(1/2) \approx 0.1476\pi$ where 
$J_{\text{A}}=J_{\text{B}}$ (see Eq. (\ref{eqn:Ja-Jb})). \\
(iii) {\em partial-AF}: 
At $\theta=\pi/2$, the system begins to have a small local magnetization 
$\VEV{\bolS_{1}{+}\bolS_{2}}$ on each rung and 
this partial magnetization orders in an AF manner. 
Note that both $\bolA$ and $\bolB$ take finite values in this phase. 
  Within the semiclassical treatment, 
the energy of the partial-AF phase is very close to that of the chirality-dominant 
phase suggesting the instability of the former against quantum fluctuations. \\
(iv) {\em F-nematic}: 
At $\theta=\cos^{-1}(-2/\sqrt{5})=\pi-\theta_{\text{sd}}\approx 0.8524\pi$ 
($t=J_{\text{bl}}=J/2+K_{4}=0$), 
the system enters a new non-magnetic phase 
which is similar to the first one (`NAF-dominant') except that 
now the $\bolA$ fields align in a {\em parallel} (ferromagnetic) manner. 
If we regard the triplet state on each rung as an effective spin-1 
state, this is nothing but the spin-nematic state 
(one can easily check $\VEV{\mathbf{T}}=\mathbf{0}$, $Q^{\alpha \beta}\neq 0$) 
where $\bolA$  plays the role of the director.  
This is why the name `F-nematic' (`F' denotes {\em ferro}) is 
used here. \\
(v) {\em ferromagnetic}: 
For $\theta > \cos ^{-1}\left(-\sqrt{\frac{32 \left(36+\sqrt{7}\right)}{1289}}\right)
\approx 0.9354 \pi$, the system is fully-occupied by the spin-1 states 
(i.e. $n_{\text{B}}=\bolA^{2}{+}\bolB^{2}=1$) 
and these spin-1s form a polarized ferromagnetic state as a whole. \\
(vi) {\em singlet-product}:  At $\theta=-\cos^{-1}(2/\sqrt{13})\approx -0.3128\pi$ 
(where $3J+2K_{4}=0$), the system becomes 
non-magnetic again through a first-order transition.  
Contrary to the ferromagnetic phase, all rungs are occupied 
by the singlet (i.e. $n_{\text{B}}=\bolA^{2}{+}\bolB^{2}=0$) in the new phase. In terms of 
the original spin-1/2 ladder model, this is nothing but the singlet-product state 
(rung-singlet).  
The original symmetry $\mathrm{SU(2)}{\times}\mathbb{Z}_{2}$ is not broken 
at all.  In fact, a simple fluctuation analysis shows that the low-energy excitation 
is the gapped triplet.  
The singlet-product phase persists until the system enters 
the `NAF-dominant' phase at 
$\theta=-\tan^{-1}(1/4)\approx -0.0780 \pi$ (i.e. $J+4K_{4}=0$), where 
the triplet gap vanishes.  

The emergence of spin-nematic phase (`F-nematic') in the absence of biquadratic interaction 
$J_{\text{bq}}$ (see Eq. (\ref{eqn:laddercoupling})) may look surprising. 
However, this can be understood within the simple mean-field argument 
presented here. 
First we note that $J_{\text{bl}}=J/2+K_{4}>0$ ($<0$) for $\theta < \pi-\theta_{\text{sd}}$ 
($\theta > \pi-\theta_{\text{sd}}$).  In the region where we have `F-nematic', 
the magnetic coupling $J_{\text{bl}}=J/2+K_{4}(<0)$ is small and 
the chirality coupling $J_{\text{B}}= K_{4}=\sin\theta$ is always positive 
(hence anti-parallel configuration 
of $\bolB$ is favored), while the NAF-coupling $J_{\text{A}}= J/2=\cos\theta/2$ is negative.  
In the `partial-AF' phase, the system tends to develop weak local moments $\mathbf{T}$ 
which align in an AF-manner to optimize the positive magnetic coupling $J_{\text{bl}}$; 
the combination $(J_{\text{A}}<0,J_{\text{B}}>0)$ stabilizes the parallel $\bolA$ and 
the anti-parallel $\bolB$ as the optimal configuration. 

In the `F-nematic' phase, on the other hand, 
the negative $J_{\text{bl}}$ naively favors ferromagnetic alignment 
of the local moments $\mathbf{T}$.  However, to realize it, configurations with 
($\bolA$-parallel, $\bolB$-parallel) or ($\bolA$-antiparallel, $\bolB$-antiparallel) 
are needed and they are inconsistent with the signs of $J_{\text{A}}$ and $J_{\text{B}}$ 
(hence frustrated).  
Since $-J_{\text{A}}>J_{\text{B}}(>0)$ and $|J_{\text{bl}}| \ll 1$,  the system lowers the energy by choosing 
a non-magnetic state (`F-nematic') 
with {\em ferromagnetic} ordering of $\bolA$ ($\bolB=\mathbf{0}$).  

If we further increase $\theta$, $J_{\text{bl}}$ gets larger while the chirality coupling 
$J_{\text{B}}$ becomes negligibly small. Then, the energy gain by forming a ferromagnetic 
(i.e. parallel $\mathbf{T}$) configuration overcomes the energy cost coming from 
the frustration in the $\bolB$-channel (i.e. parallel-$\bolB$ for positive $J_{\text{B}}$) 
and hence the ferromagnetic state is stabilized. 

So far, we have presented the mean-field description of the phases of 
the spin-1 boson model and found that in some of these phases, 
rotational symmetry is spontaneously broken at the mean-field level.  
However,  in one dimension, strong quantum fluctuations 
may destroy the ordered states predicted by the mean-field theory.  
In fact, as we will show in the next section by using effective field theories, 
some of the ordered phases 
are replaced by gapped short-range phases which do not break rotational symmetry.  
\section{Low-energy effective actions}
\label{sec:effective-action}
Having established the mean-field phase diagram, 
we now take into account quantum fluctuations in a semiclassical fashion.  
From the viewpoint of the competition between spin and chirality, 
it is interesting to develop the continuum descriptions of the two phases 
--the NAF-dominant phase and the chirality-dominant phase. 
\subsection{Low-Energy Fluctuations}
\label{sec:low-energy-fluc}
\begin{figure}[tbh]
\begin{center}
\includegraphics[scale=0.46]{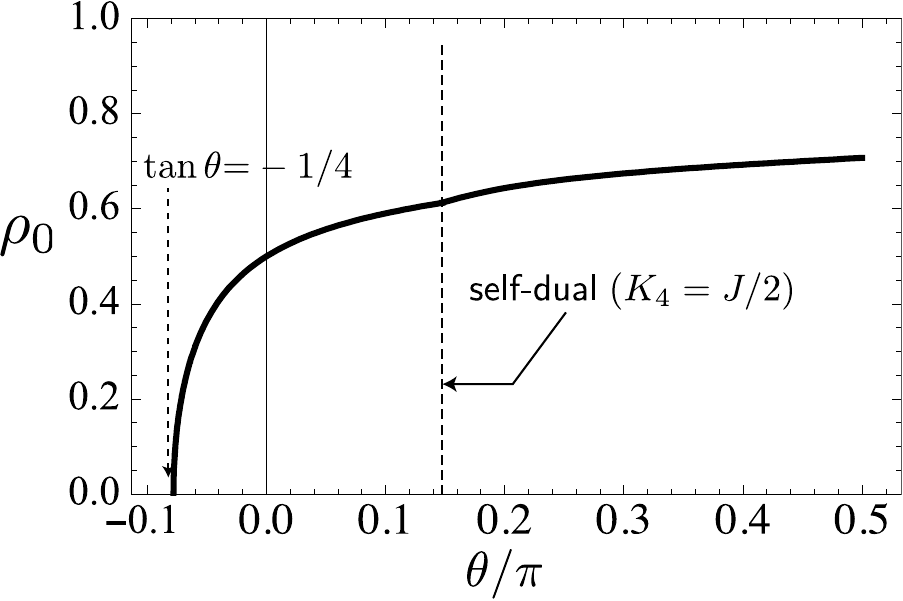}
\includegraphics[scale=0.46]{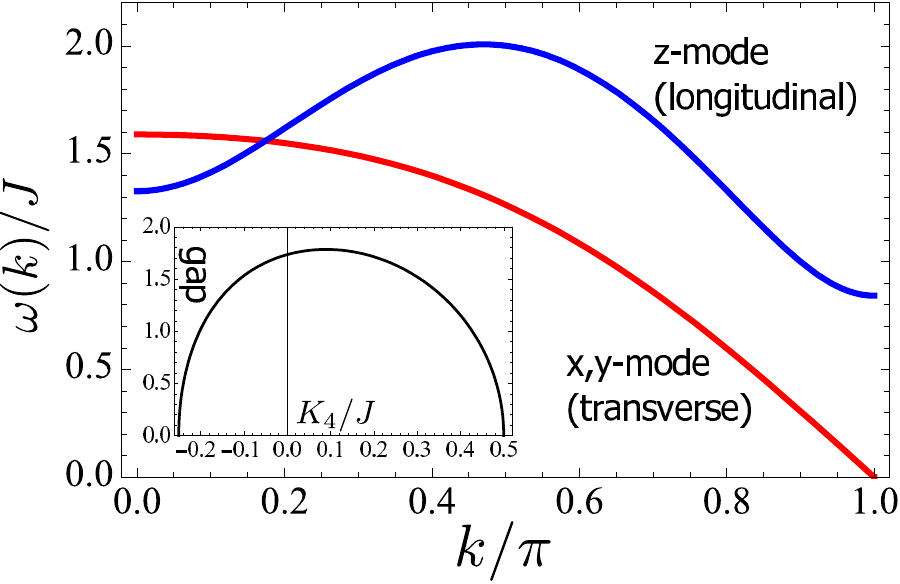}
\caption{(color online) 
(Left): Square-root $\rho_{0}$ (Eq. (\ref{eqn:rho_0})) of the boson density $n^{\text{B}}$ 
as a function of $K_{4}/J$.  Note that the boson density $\rho_{0}^{2}$ 
is always finite.  
(Right): The low-energy dispersion obtained by a semiclassical 
(spinwave-like) calculation ($K_{4}/J=0.45$).  Inset shows 
the variation of the $z$-mode gap (at $k=\pi$) as a function 
of $K_{4}/J$. 
The staggered ordering of $\bolA$ 
in the $z$-direction is assumed.  Among the three modes ($x$, $y$ and 
$z$), the longitudinal one ($z$-direction) may be regarded as 
the density fluctuation and is massive, while the remaining two are 
gapless (Goldstone modes) corresponding to the symmetry breaking: 
O(3)$\rightarrow$O(2) (In 1D, these Goldstone modes eventually get gapped by 
strong quantum fluctuations).    Only at the self-dual point $K_{4}/J=1/2$, 
the longitudinal branch ($z$-mode) shows a gapless $k$-linear behavior. 
\label{fig:three-modes}}
\end{center}
\end{figure}

In constructing the continuum limit, it is important to find out 
the relevant low-energy degrees of freedom.  
To this end, let us calculate the low-energy spectrum 
in the semiclassical approximation.  
First we parametrize the semi-classical configuration as:
\begin{subequations}
\begin{equation}
\begin{split}
& \Vec{A}^{(0)}_{r} = (0,0,(-1)^{r}\rho_{0}) \; , \;\;
\Vec{B}^{(0)}_{r}=(0,0,0) \\
& \qquad 
\text{for } -\tan^{-1}(1/4) \leq \theta < \theta_{\text{sd}}=0.1476\pi
\end{split}
\label{eqn:classical-GS}
\end{equation}
or
\begin{equation}
\begin{split}
& \Vec{A}^{(0)}_{r} = (0,0,0) \; , \;\;
\Vec{B}^{(0)}_{r}=(0,0,(-1)^{r}\rho_{0}) \\ 
&\qquad 
\text{for }\theta_{\text{sd}} 
< \theta < \pi/2 
 \; ,
\end{split}
\label{eqn:classical-GS1}
\end{equation}
\end{subequations}
where $\rho_{0}\equiv \sqrt{n_{\text{B}}}$  
$(0\leq \rho_{0} \leq 1)$ and is given by (see Appendix 
\ref{sec:App-continuum-limit}):
\begin{equation} 
\rho_{0} = 
\begin{cases}
\frac{1}{2}\sqrt{\frac{J+4K_{4}}{J+2K_{4}}}  \quad & \text{for }
K_{4} < J/2 \\
\frac{1}{4}\sqrt{\frac{-J+8K_{4}}{K_{4}}} 
\quad & \text{for } K_{4} > J/2 \; .
\end{cases}
\label{eqn:rho_0}
\end{equation}
The value of $\rho_{0}$ is plotted in the left panel of 
Fig.~\ref{fig:three-modes}. 

Now that we have determined the classical (or mean-field) ground state, 
we are at the point of calculating the spectrum of the low-energy 
excitations.  
Let us consider the following small 
deviation $\delta\bolA$, $\delta\bolB$ from the classical ground 
state:
\begin{equation}
\Vec{A}_{r} = \Vec{A}^{(0)}_{r} + \delta\Vec{A}_{r} \; , \;\;
\Vec{B}_{r} = \Vec{B}^{(0)}_{r} + \delta\Vec{B}_{r} \; .
\label{deviation}
\end{equation}
If we plug the above expressions into the classical equation of 
motion and retain terms up to 1st-order in $\delta \bolA$ and 
$\delta \bolB$, we obtain the following `spin-wave' spectra:
\begin{widetext}
\begin{subequations}
\begin{align}
& \omega_{x}(k) = \omega_{y}(k) 
 =\frac{J \cos \left(\frac{k}{2}\right) \sqrt{(3 J+4 K_{4})
 (3 J+4 K_{4} - J\cos (k))}}{\sqrt{2} (J+2 K_{4})}  
\\
& \omega_{z}(k)= 
 \sqrt{\frac{\cos (k) J^3+4 (J+2 K_{4}) J^2
+2 K_{4} \left(J^2-8 K_{4} J-16
   K_{4}^2\right) \cos ^2(k)}{J+2 K_{4}}} 
\; .
\end{align}
\end{subequations}
\end{widetext}
It is easy to verify that the transverse modes ($\omega_{x,y}(k)$) 
have a linear dispersion near the zone boundary $k=\pi$:
\begin{equation}
\omega_{x,y}(k) \approx 
\frac{J \sqrt{(J+K_{4})(3J+4K_{4})}}{\sqrt{2}
 (J+2 K_{4})}  |k-\pi| \; , 
 \label{eqn:linearized-SW}
\end{equation}
while the longitudinal one ($\omega_{z}$) is gapped 
(unless $K_{4}=J/2$).   
These linearly dispersive modes roughly describe transverse 
fluctuations of the director vectors $\bolA$ and similar 
to the usual AF magnons.  
The spectrum for the $\bolB$-ordered phase (`chirality-dominant' phase), which is 
realized when $K_{4}\gg J$, can be obtained in a similar fashion.   
\subsection{Continuum Limit}
\label{sec:continuum-limit}
The above analysis suggests that we should keep 
the low-energy order parameter field around $k=\pi$ in constructing the continuum limit 
for the dominant-NAF phase. 
Specifically, we use the following {\em two-sublattice} (plaquette-wise, actually) 
ansatz to parametrize 
the $\bolA$ and $\bolB$ fields:
\begin{equation}
\begin{split}
&\bolA_{r}+i\, \bolB_{r} = 
\sqrt{\rho_{0}^{2}+a\, l_{\phi}(x)} \, 
(-1)^{r}\be^{i\phi(x)} \\
& \quad \times 
\left\{\sqrt{1-a^{2} \mathbf{l}^{2}_{\Omega}(x)}\, \bolOmega(x) + 
i a(\bolOmega{\times}\mathbf{l}_{\Omega}) \right\}  \; ,
\end{split}
\label{eqn:AB-by-Omega}  
\end{equation}
where the order parameter field ({\em director}) satisfies $|\bolOmega(x)|=1$ 
and $\bolOmega \cdot \mathbf{l}_{\Omega}= 0$.  
Physically, the two small fields 
$\mathbf{l}_{\Omega}$ and $l_{\phi}$
 are respectively  the generators of spin-SU(2) and charge-U(1):
\begin{equation}
\begin{split}
& \mathbf{T}^{\alpha}_{r} = 
-i\epsilon_{\alpha \beta \gamma}b_{\beta}^{\dagger}b_{\gamma} = 2\rho_{0}^{2}a\,(\mathbf{l}_{\Omega})^{\alpha}+ \cdots \\
& n_{\text{B}}
=\sum_{\alpha}b^{\dagger}_{\alpha}b_{\alpha} = \rho_{0}^{2} + a\, l_{\phi}(x)  + \cdots \; .
\end{split}
\end{equation}

The vector order parameter $\bolOmega$ represents spin and chirality 
in a unifying manner. 
When the phase $\phi$ is dominantly around $\phi=0$ or $\pi$, 
the order parameter $\bolOmega$ describes the transverse fluctuations of 
the $(\pi,\pi)$ antiferromagnetic order represented by $\bolA$: 
\begin{equation}
\bolOmega \sim(\Vec{S}_{1}{-}\Vec{S}_{2})_{q=\pi} \; .  
\end{equation}  
When $\phi$ fluctuates around $\pm\pi/2$, on the other hand, 
$\bolOmega$ constitutes the imaginary part of the quantity $\bolA+i\bolB$ and 
describes the fluctuations of (the staggered component of) 
the vector chirality $\bolB$:
\begin{equation}
\bolOmega \sim(\Vec{S}_{1}{\times}\Vec{S}_{2})_{q=\pi}  \; .  
\end{equation}

As in the usual Haldane mapping\cite{Auerbach-book}, in order to write down 
the effective action, 
we assume that the two fields $\mathbf{l}_{\Omega}$ and $l_{\phi}$ are small 
compared with $\bolOmega$ and retain terms up to second-order 
in the calculation.   
The condensate amplitude $\rho_{0}$ and the phase angle $\phi_{0}$ 
of the ground state can be determined by minimizing the constant part of the effective action:
\begin{eqnarray}
E_{0} &=& \rho_0^2 \left\{ \rho_0^2 (J+6 K_4 ) {-} 6 K_4\right\}
\\ \nonumber
&-& \rho_0^2 (1-\rho_0^2)  (J-2 K_4) \cos (2 \phi_{0})  \; .
\label{eqn:eff-Ham-constant}
\end{eqnarray}
Depending on the sign of $(J-2K_{4})$, either $\phi_{0}=0,\pi$ or 
$\phi_{0}=\pi/2,3\pi/2$ is chosen. 

After some algebra, we obtain (see Appendix \ref{sec:App-continuum-limit}
for details): 
\begin{equation}
\begin{split}
{\cal S}_{\text{eff}}[\bolOmega] =& 
\frac{1}{2}\chi_{\text{s}} \int\!dx dt \,
(\partial_{t}\bolOmega)^{2} - \frac{1}{2}\rho_{\text{s}}
\int\!dx dt \, (\partial_{x} \bolOmega)^{2}  \\
& + \frac{1}{2}\chi_{\text{c}} \int\!dx dt \,
(\partial_{t}\phi)^{2} 
- \frac{1}{2}\rho_{\text{c}}
\int\!dx dt \, (\partial_{x} \phi)^{2} \\
&\quad + g(J,K_{4})
 \int\!dx dt \,
\cos (2 \phi (x,t))\; ,
\end{split}
\label{eqn:eff-Ham}
\end{equation}
where the explicit expressions for the five coupling constants (spin/charge susceptibility 
$\chi_{\text{s/c}}$, the spin/charge stiffness $\rho_{\text{s/c}}$ and 
the sine-Gordon coupling $g(J,K_{4})$) are given 
in Appendix (Eq.~(\ref{eqn:chi-rho-A})). 
Note that a similar effective action (without the cosine-term) has been derived 
in the context of $F=1$ spinor Bose-Einstein condensate.
\cite{Demler-Z-02,*Zhou-review-03,Essler-S-T-09}

The spin part of the effective action is nothing but the O(3) non-linear sigma model 
in $(1{+}1)$-dimensions, where the long-range-ordered ground state with 
the gapless $k$-linear Goldstone modes is replaced by a quantum-disordered one with 
the gapped triplet excitations ({\em triplon}).\cite{Zamolodchikov-Z-79}
Since near $\theta=0$ and $\pi$, $\bolOmega$ may be thought of as 
the order parameter for the standard $(\pi,\pi)$ AF ordering, 
the spin part describes the short-range antiferromagnetic fluctuations for 
$J>K_{4}$.  Since $\bolA$ and $\bolB$ play the role of 
the director in the spin-nematic, one may think that the order parameter manifold is 
$\text{RP}^{2}$.  However, in contrast to the case of the usual spin nematic, the directors 
($\bolA$ and $\bolB$) themselves are directly related to the physical observables 
(i.e. $\bolS_{1}{-}\bolS_{2}$ and $\bolS_{1}{\times}\bolS_{2}$, respectively) and 
the order parameter in fact is defined on the two-dimensional sphere $S^{2}$.   
In principle, there can be a topological term associated with $\Pi_{2}(S^{2})$ 
in the above effective action.  
However, the above direct mapping suggests that there is no such topological term 
in the effective action.  

On top of the spin part, there is the spin-chirality (or, in terms of spin-1 bosons, 
`charge') part which 
describes the dynamics of the Bose phase $\phi$;   
the spin-chirality part of the effective action is given by the sine-Gordon model 
with the $(J,K_{4})$-dependent couplings.  Depending on the sign of 
$g(J,K_{4})$, the cosine term pins the ground-state phase either at $\phi=0,\pi$ 
(for $g(J,K_{4})>0$) or at $\phi=\pi/2,3\pi/2$ (for $g(J,K_{4})<0$).  
Since $\phi \mapsto \phi+\pi$ amounts to the 1-site translation 
$r\mapsto r+1$ (see Eq.~(\ref{eqn:AB-by-Omega})), 
one can see that in both cases there appear two-fold degenerate ground states related  
by 1-site translation; the sine-Gordon soliton connects the two degenerate 
ground states.  At the self-dual point $K_{4}=J/2$, the sine-Gordon coupling disappears 
and the charge (i.e. spin-chirality) part reduces to the Tomonaga-Luttinger model 
as is expected from the previous results.\cite{Lecheminant-T-05,Lecheminant-T-06-SU4}

Now we proceed to a more interesting case of the chirality-dominant phase 
realized in the large-$K_{4}$ region. In this phase, $\phi_{0}$ is locked at 
$\phi=\pm \pi/2$ and the effective action (\ref{eqn:eff-Ham}) 
describes the short-range (staggered) fluctuations in the chirality channel; 
the staggered component of the $\bolB$-field plays a role of 
the order-parameter field: $\bolOmega \sim(\Vec{S}_{1}{\times}\Vec{S}_{2})_{q=\pi}$.  
As before, the ground state is quantum-disordered with  
a gapped triplet excitation, which is created not by the spin operators themselves but by 
the staggered component of the vector chirality 
$(\Vec{S}_{1}{\times}\Vec{S}_{2})_{q=\pi}$.  This agrees with the results 
of numerical simulations.\cite{Lauchli-S-T-03,Hikihara-M-H-03} 
Since the duality transformation ($\varphi=\pi/2$ in Eq.~(\ref{eqn:spin-chirality-by-AB}) 
or $\phi \mapsto \phi +\pi/2$ in Eq.~(\ref{eqn:AB-by-Omega})) 
interchanges two vector fields $\bolA$ (`spin') and $\bolB$ (`chirality'), 
these two effective action (\ref{eqn:eff-Ham}) 
clearly exhibit the `dual' nature of spin and 
chirality in the model (\ref{eqn:ladder}).

To summarize, we have obtained the O(3) non-linear sigma model   
for the spin dynamics in the NAF/chirality-dominant phases;  
deep inside these phases, 
where the spin-chirality ($\theta$) fluctuations have a large gap and 
are well separated from the transverse spin fluctuations 
(see the inset of Fig.~\ref{fig:three-modes}),  
we may expect that the {\em pure} sigma model provides us with 
a good description of the low-energy 
physics far away from the self-dual point $K_{4}/J=1/2$.  
Since a similar non-linear sigma model is obtained 
in the standard field-theory treatment\cite{Senechal-95,*Sierra-96} 
of the two-leg ladder with $K_{4}=0$, one may identify the NAF-dominant phase 
with the rung-singlet (RS) phase.  However, as we will see below, this is not the case. 
It is important to note that the above derivation of the effective action  
applies to higher dimensional cases as well with due modification 
of their coupling constants $\rho_{\text{s,c}}$ and $\chi_{\text{s,c}}$. 

When the self-dual point $K_{4}/J=1/2$ is approached, on the other hand, 
the gap of the $\omega_{z}$-branch (i.e. the $\phi$-channel) decreases and 
eventually becomes smaller than the dynamically-generated spin gap; 
exactly at the point, the longitudinal mode exhibits 
the gapless $k$-linear behavior%
\footnote{The explicit form of the normal modes around $k=\pi$ shows 
that the corresponding fluctuation is of the form 
$\bolA \mapsto \be^{i\delta \phi}\bolA$.} around $k=\pi$ which we may identify 
with the gapless spin-chirality (i.e. the U(1)-phase of the spin-1 hardcore boson) 
fluctuations represented by the Tomonaga-Luttinger model  also found in the bosonization 
analysis.\cite{Lecheminant-T-05,Lecheminant-T-06-SU4}  
As the gap in the spin sector (which is described by the non-linear sigma model) 
is always finite, the quantum phase transition between NAF-dominant phase 
and the chirality-dominant one is described by the $c=1$ Tomonaga-Luttinger 
liquid.\cite{Lecheminant-T-06-SU4} 
In fact, around the self-dual point, the spin-singlet phase (`charge') fluctuations 
play the primary role.  

Away from the self-dual point, 
the anomalous hopping term generates the $\cos 2\phi$-term 
which pins the ground state at $\phi=0,\pi$ 
or at $\phi=\pi/2,3\pi/2$ to stabilize the two-fold degenerate ground states
of the model (\ref{eqn:ladder}) as far as the condensate amplitude $\rho_{0}$ is finite; 
the two-fold degeneracy corresponds to breaking of the $\mathbb{Z}_{2}$-symmetry 
$(\bolA,\bolB)\mapsto (-\bolA,-\bolB)$ (i.e. the interchange of the two chains 
$1\leftrightarrow 2$) or the 1-site translation $r \mapsto r+1$.  
In 1D, the rotation symmetry is restored and the only broken symmetry is 
the $\mathbb{Z}_{2}$. 
This is what we have in the staggered-dimer phase and 
the staggered-scalar-chirality phase.\cite{Lauchli-S-T-03} 
In the case of the two-leg ladder, the spin excitations are always the gapped triplons 
in the three main phases ({\em singlet-product}, {\em NAF/chirality-dominant}) 
in the $J>0$ region (see Fig. \ref{fig:classical-AB-phases}) 
and only the presence/absence of the $\mathbb{Z}_{2}$-symmetry distinguishes 
between the first- and the latter two phases.    
In this respect,   
the NAF-dominant- and the chirality-dominant phase may be identified respectively 
with the staggered dimer phase and the staggered scalar chiral phase 
found in Ref.~\onlinecite{Lauchli-S-T-03}.  
Since our semiclassical treatment underestimates the effect of quantum fluctuations 
brought about by the pairing term, the $\mathbb{Z}_{2}$-broken ordered states 
(NAF/chirality-dominant phases) are more statbilized than in the actual two-leg 
ladder.\cite{Lauchli-S-T-03}

Last, we briefly touch upon the effective action in the F-nematic phase. 
A similar fluctuation analysis shows that the gapless $k$-linear modes appear 
at $k=0$ in this case. 
The calculation is essentially the same as in the previous two cases except that 
we adopt the following parametrization reflecting the gapless transverse 
modes at $k=0$:
\begin{equation}
\begin{split}
&\bolA_{r}+i\, \bolB_{r} = 
\sqrt{\rho_{0}^{2}+a\, l_{\phi}(x)} \, 
\be^{(-1)^{r}i\phi(x)} \\
& \quad \times 
\left\{\sqrt{1-a^{2} \mathbf{l}^{2}_{\Omega}(x)}\, \bolOmega(x) + 
i a(\bolOmega{\times}\mathbf{l}_{\Omega}) \right\}  \; .
\end{split}
\label{eqn:eqn:ansatz-F-nem}  
\end{equation}
The staggered definition of the Bose phase $\be^{(-1)^{r}i\phi(x)}$ 
has been inspired by the gapless $k$-linear mode appearing at 
the point $J+2K_{4}=0$ reflecting the existence of an alternating 
spin-chirality U(1)-symmetry.\cite{Hikihara-Y-08}  
As in the above cases, we carry out gradient expansion and the subsequent 
Gaussian integration over $\mathbf{l}_{\Omega}$, $l_{\phi}$ to obtain 
the non-linear sigma model (\ref{eqn:eff-Ham}) with 
different coupling constants (see Eq. (\ref{eqn:chi-rho-F-NEM})).  
At the transition point ($J+2K_{4}=0$) between partial-AF and F-nematic, 
the sine-Gordon interaction vanishes and the transition is described by 
the Tomonaga-Luttinger liquid.  
Again, quantum fluctuations open a gap in the spin sector and 
the ground state is characterized by the short-range correlation 
of the collinear order parameter $\bolOmega \sim \bolS_{1}{-}\bolS_{2}$ at 
$k=0$, which is consistent with the numerical observation.\cite{Lauchli-S-T-03}
\section{Effects of Magnetic Field and Mutual Induction Phenomena}
\label{sec:effect-magnetic-field}
In this section, we investigate the effect of a magnetic field 
on the two-leg spin ladder with a ring-exchange  by means of our semiclassical approach.
We show, as a by-product, that the spin-chirality 
duality leads to an interesting phenomenon--{\em mutual induction 
between spin and chirality} degrees of freedom.  

To this end, it is convenient  to enlarge the parameter space of model (\ref{eqn:ladder}) and consider  a two-leg ladder 
with a general four-spin exchange interaction: \cite{Lecheminant-T-06-SU4}
\begin{equation}
{\cal H} = g_{1}{\cal H}_{1}+ g_{2}{\cal H}_{2}+ g_{3}{\cal H}_{3}
+g_{4}{\cal H}_{4} +g_{5}{\cal H}_{5} + g_{6}{\cal H}_{6} \; ,
\label{eqn:general-ladder}
\end{equation}
where the six building blocks are given as:\cite{Lecheminant-T-06-SU4}
\begin{subequations}
\begin{align}
\begin{split}
{\cal H}_{1} = \sum_{r}
(\bolS_{1,r}+\bolS_{2,r}){\cdot}(\bolS_{1,r+1}+\bolS_{2,r+1}) 
\end{split}
\label{eqn:int_1}
\\
\begin{split}
{\cal H}_{2} = \sum_{r} (\bolS_{1,r}-\bolS_{2,r}){\cdot}(\bolS_{1,r+1}-\bolS_{2,r+1}) 
\end{split}
\label{eqn:int_2}
\end{align}
\begin{align}
\begin{split}
{\cal H}_{3}
&= 4 \sum_{r} \bigl[
(\bolS_{1,r}{\cdot}\bolS_{1,r+1})(\bolS_{2,r}{\cdot}\bolS_{2,r+1}) \\
& \phantom{4 \sum_{r} \bigl[} 
+(\bolS_{1,r}{\cdot}\bolS_{2,r+1})(\bolS_{2,r}{\cdot}\bolS_{1,r+1})
\bigr]
\end{split}
\label{eqn:int_3}
\\
\begin{split}
{\cal H}_{4} &= 4
\sum_{r} (\bolS_{1,r} {\times} \bolS_{2,r}) {\cdot} (\bolS_{1,r+1} {\times} \bolS_{2,r+1}) 
\end{split}
\label{eqn:int_4}
\end{align}
\begin{align}
{\cal H}_{5} &= \frac{1}{2}\sum_{r} (\bolS_{1,r}{\cdot}\bolS_{2,r}+
\bolS_{1,r+1}{\cdot}\bolS_{2,r+1})
\label{eqn:int_6}
\\
{\cal H}_{6} &= \sum_{r} (\bolS_{1,r}{\cdot}\bolS_{2,r})
(\bolS_{1,r+1}{\cdot}\bolS_{2,r+1}) \; .
\label{eqn:int_7}
\end{align}
\end{subequations}
The six coupling constants are expressed in terms of the six bosonic ones as:
\begin{equation}
\begin{split}
& g_1=J_{\text{bl}}-J_{\text{bq}}/2 \, ,\; g_2=(t+u)/2\, , \; g_3=J_{\text{bq}}/2 \, , \\
& g_4=(t-u)/2 \, , \; g_5= 2J_{\text{bq}} -\mu+3V_{\text{c}}/2\, , \; 
g_6=V_{\text{c}} \; .
\end{split}
\end{equation}
Under the spin-chirality duality symmetry (\ref{spinchirality}), the coupling
constants of model (\ref{eqn:general-ladder}) transform as:
$g_{2} \leftrightarrow g_{4}$, while all the others remain invariant. 
\subsection{Mean-field analysis}
\begin{figure}[htb]
\begin{center}
\includegraphics[scale=0.5]{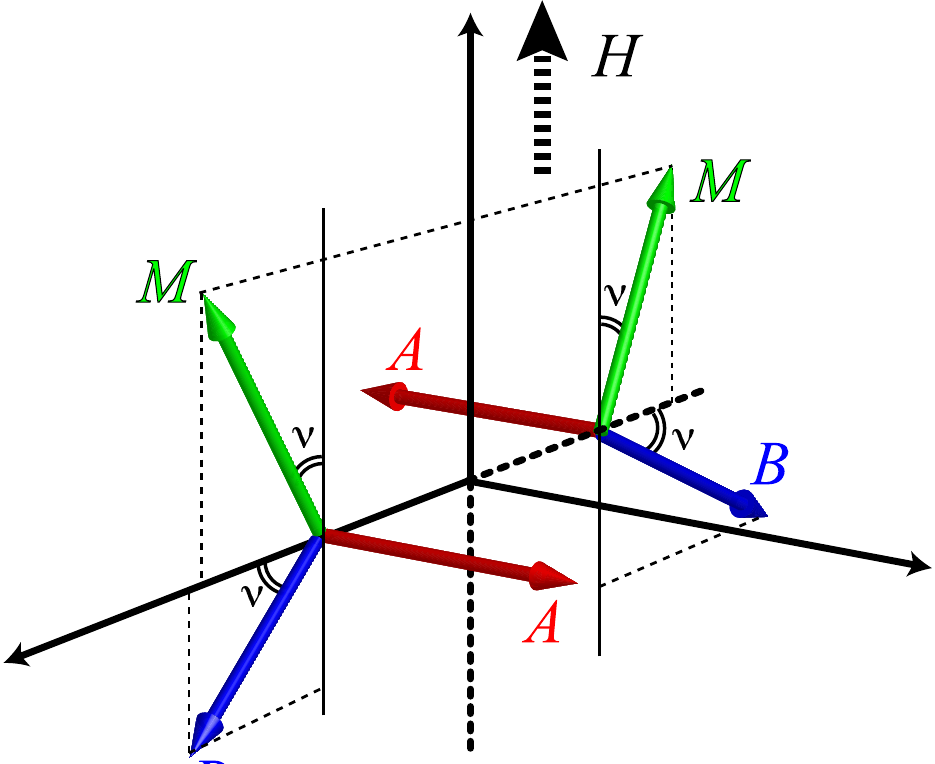}
\caption{(color online) A naively expected high-field configuration of 
$\bolA \propto \bolS_{1}{-}\bolS_{2}$, 
$\bolB \propto \bolS_{1}{\times}\bolS_{2}$, and 
$\VEV{\mathbf{T}}\propto \bolS_{1}{+}\bolS_{2}$.  
Spin ($\mathbf{T}$) canting induces a {\em uniform} $z$-component 
$(\bolS_{1}{\times}\bolS_{2})^{z}$.  
If a similar spin-canted state is realized for large enough 
$g_{4}$, we may expect a state where local magnetization is induced 
asymmetrically in the two chains 
$\VEV{S^{z}_{1}}\neq \VEV{S^{z}_{2}}$. 
\label{fig:canted-AB}}
\end{center}
\end{figure}

Let us consider the situation where the magnetic interaction $J_{\text{bl}}$ 
of model  (\ref{eqn:spin1tJ})  
(i.e.  $g_{1}$ in the ladder model (\ref{eqn:general-ladder}))   
is dominant and the interaction $g_{2}(>0)$ in 
the spin-spin channel is stronger than that in the chirality channel $g_{4}$.  
The idea is as follows. 
When a sufficiently strong magnetic 
field (say, in the $z$-direction) 
induces uniform magnetization, the magnetic moments 
${\bf M}=\bolS_{1}{+}\bolS_{2}=-2\bolA{\times}\bolB$ (see Eq. (\ref{eqn:local-moment-by-AB}))
assume (at least in the semiclassical sense) the {\em canted} configuration 
as is shown in Fig.~\ref{fig:canted-AB}.  
A simple mean-field argument given above concludes that the anti-parallel 
ordering of $\bolA \propto \bolS_{1}{-}\bolS_{2}$ forces 
the canting of the chirality vector $\bolB \propto \bolS_{1} {\times} \bolS_{2}$ leading to the appearance of 
finite vector chirality in the field direction: \footnote{%
In 1D, the projection of $\bolB$ onto the plane perpendicular to 
$H$ cannot order and only the $z$-component takes a finite 
expectation value; in higher dimension, on the other hand, it is possible 
that both $\bolA$ and $\bolB$ exhibit long-range order.} 
$(\bolS_{1}{\times}\bolS_{2})^{z} \ne 0$ 
(see Fig.~\ref{fig:canted-AB}).  
Note that the mean-field (or variational) energy is invariant under the {\em global} 
$\mathbb{Z}_{2}$-symmetry: $\bolA\mapsto -\bolA$, $\bolB\mapsto -\bolB$ 
corresponding to the interchange of the two chains $1\leftrightarrow 2$. 
Therefore, the appearance of uniform vector chirality in the low-field phase is accompanied by 
the $\mathbb{Z}_{2}$-symmetry breaking. 

According to the spin-chirality duality (\ref{spinchirality}), 
the same argument applies to the case where the chirality channel $g_{4}(>0)$ is 
dominant; 
in this case, the ordering in the chirality channel induces  
finite $(\bolS_{1}{-}\bolS_{2})^{z}$, that is, magnetization is induced 
{\em asymmetrically} in the upper- and the lower chain.  
It is important to note that, when the interaction in the spin-spin 
channel is dominant, the chirality is induced and vice versa. 
As is clear from the above argument, this is not restricted to 
one dimension; our semiclassical approach immediately predicts that 
in two dimensions too.  One of the most important conclusions 
of our bosonic  effective theory is that 
this {\em mutual induction} is an observable consequence 
of the spin-chirality duality.   

\begin{figure}[ht]
\begin{center}
\includegraphics[scale=0.4]{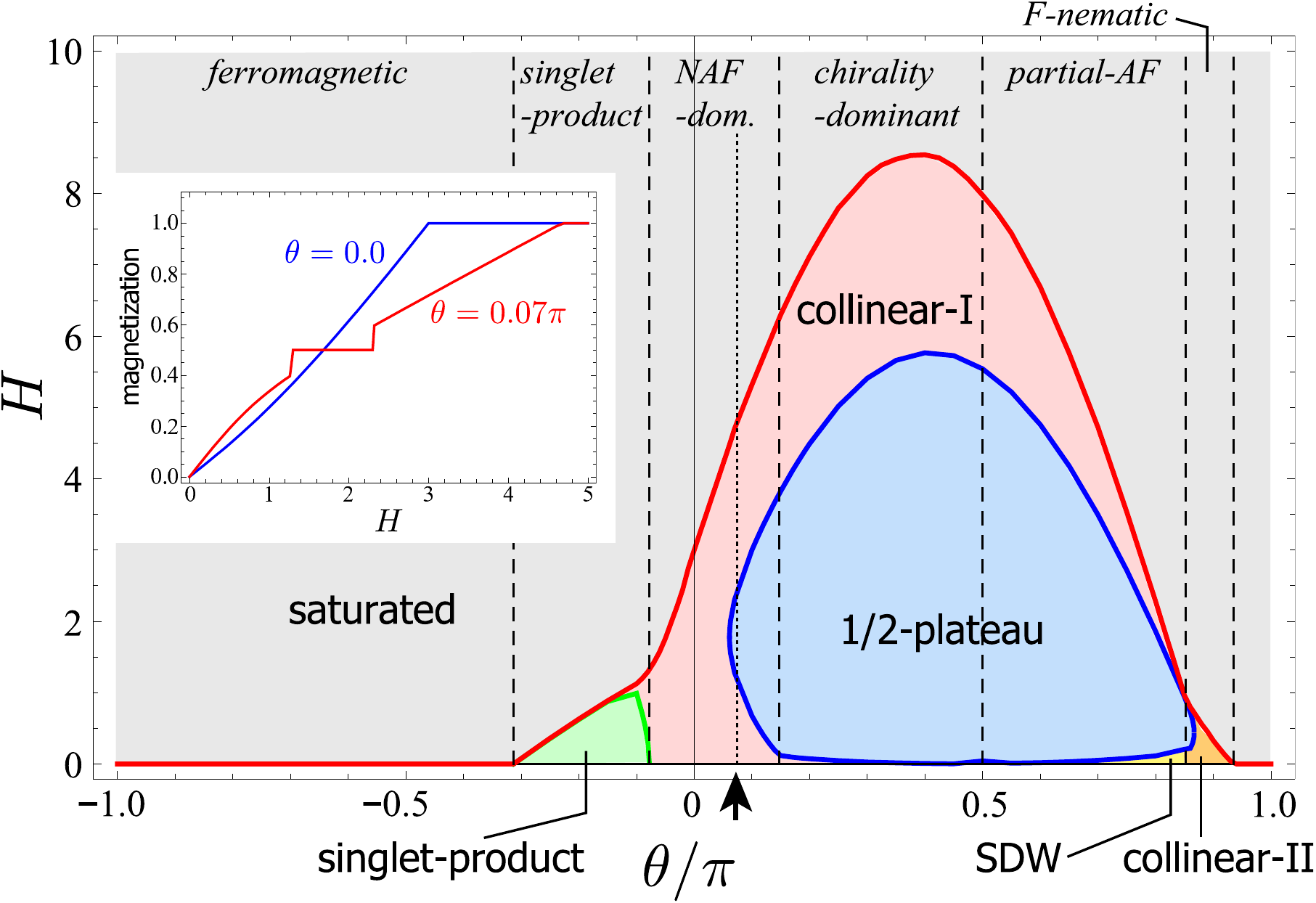}
\caption{(color online) Mean-field phase diagram of the ring-exchange ladder 
in magnetic field ($J=\cos\theta$, $K_{4}=\sin\theta$). 
The large portion of the parameter space is occupied by 
1/2-plateau phase.  All the phases except the singlet-product phase are 
{\em collinear} and are characterized by parallel or anti-parallel alignment of 
$\bolA$, $\bolB$ and $\VEV{\hat{\mathbf{T}}}$.  The canted phase anticipated 
in the text does {\em not} appear in this parameter space. 
Inset: magnetization curves for $\theta=0$ and $0.07\pi$ (marked by an arrow head 
in the main panel). 
\label{fig:ring-ladder-PD-H}}
\end{center}
\end{figure}

It would be interesting to check if the simplest ring-exchange ladder 
(\ref{eqn:ladder}) really supports this phase or not.  
Again we can use mean-field analysis to map out the phase diagram 
in the presence of magnetic field. The result is summarized 
in Fig.~\ref{fig:ring-ladder-PD-H}.  Except for the trivial saturated phase 
which covers the entire ferromagnetic phase as well as the high-field region, 
the 1/2-plateau phase, which is characterized by spin-polarized spin-1 
bosons sitting every other site, occupies the large portion of the phase 
diagram.  This is consistent with the previous results obtained by 
numerical simulations and strong-coupling 
expansions.\cite{Sakai-H-99,*Nakasu-T-H-O-S-01,Hikihara-Y-08}  

\begin{figure}[H]
\begin{center}
\includegraphics[scale=0.5]{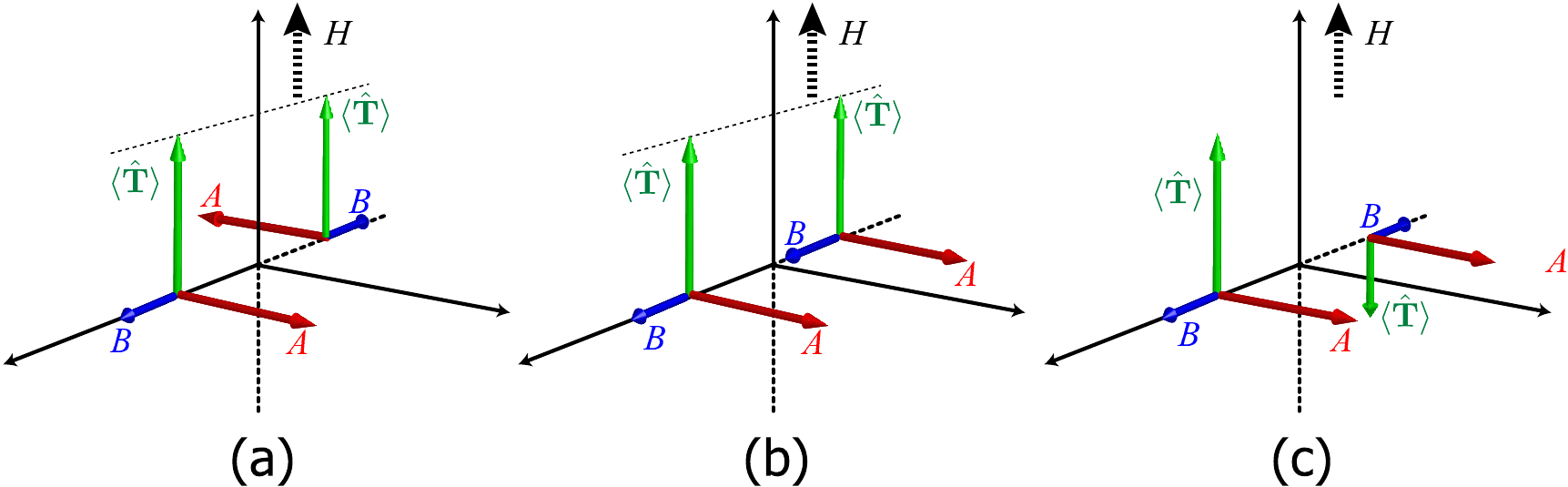}
\caption{(color online) Three collinear configurations exhibiting smooth 
magnetization process: (a) `collinear-I', (b) `collinear-II' and (c) `SDW'. 
The configuration (a) generically appears in the high-field region, while 
(b) realizes when the field is applied in the `F-nematic' phase (see Fig.~\ref{fig:ring-ladder-PD-H}). 
In the configurations (a) and (b), $z$-component of local magnetization $T^{z}$ and 
local boson density $n_{\text{B}}$ are uniform. 
In the (c) configuration, on the other hand, these quantities alternate along the ladder to 
form a density-wave state. 
\label{fig:spin-config-H}}
\end{center}
\end{figure}

Magnetization increases 
smoothly in the two collinear phases ({\em collinear-I} and {\em collinear-II}). 
In the collinear-I phase, both $\bolA$ and $\bolB$ align in an AF
manner and are perpendicular to the field (hence the local moments 
$\propto \bolA{\times}\bolB$ are parallel to the field. See Fig.~\ref{fig:spin-config-H}(a)).  
It exists typically in the low-field 
region of `NAF-dominant' phase and in the high-field region near saturation. 
In fact, it is easy to show, by finding the exact single-magnon wave function, 
that it is a generalizations of 
(the simplest version of) the triplon condensate at momentum $k=\pi$ 
which is used frequently in analyzing the magnetization process of dimer systems; 
here not only the spin component parallel (i.e. $T^{z}=+1$) to the field 
but also the anti-parallel (i.e. $T^{z}=-1$) one are taken into account.%
\cite{Romhanyi-T-P-11} 
The collinear-II phase (Fig.~\ref{fig:spin-config-H}(b)) is similar to the collinear-I except that 
here both $\bolA$ and $\bolB$ align ferromagnetically. 
In this sense, this is thought of as the classical counterpart of 
the triplon condensate at $k=0$.  Typically, it appears 
when the field is applied in the `F-nematic' phase.  

On top of them, yet another collinear phase (dubbed {\em SDW} 
for spin density wave) appears 
in the `partial-AF' phase when the field is very weak 
(a tiny region below the 1/2-plateau). In the SDW phase, the local moments $\VEV{\hat{\mathbf{T}}}$ 
form an up-down pattern along the field direction while the lengths of the moments on 
different sublattices are not equal (Fig.~\ref{fig:spin-config-H}(c)).  

Our mean-field results basically agree with those of density-matrix-renormalization-group~\cite{White-92}  (DMRG) 
simulations;\cite{Hikihara-Y-08} the exceptions are the `SDW' phase and 
`collinear-II' near saturation. In fact, in the region where we found `SDW', 
 DMRG simulations observed a phase with the dominant vector-chiral correlation 
(in the transverse direction).  As we have remarked above, already at the mean-field 
level, the energies of `chirality-dominant' phase and `SDW' are very close to each 
other and the correct treatment of quantum fluctuations might stabilize 
the chirality over `SDW' state.  `Collinear-II' state in the classical limit translates 
in the quantum description 
to the {\em single-boson} 
(either triplons over the rung singlet or singlet-rungs in the spin-polarized state) 
condensate at $k=0$.  In the low-field region, this is consistent with 
the phase diagram (Fig.~1 and 5 of Ref.~\onlinecite{Hikihara-Y-08}) obtained numerically. 
However, what has been observed near saturation is the phase characterized by 
the condensation of magnon bound states. The failure of our mean-field theory 
in describing this magnon-pair phase is not surprising since our wave function is tailored 
to describe the single-boson condensate. 

Clearly, the mean-field $(\theta,H)$-phase diagram 
(Fig.~\ref{fig:ring-ladder-PD-H}) of the ring-exchange ladder (\ref{eqn:ladder})
shows that the spin-canted state illustrated in Fig.~\ref{fig:canted-AB}
is not realized at least in the simplest ladder.  
In fact, after this phenomenon of magnetization-induced vector chirality 
has been predicted first in one dimension 
by means of bosonization,\cite{Sato-07} 
extensive DMRG study\cite{Hikihara-Y-08} 
has been carried out to show that the pure ring-exchange model (\ref{eqn:ladder}) in a magnetic field
is not sufficient to support the predicted phase with finite uniform (vector) chirality.  
This is consistent with our mean-field study. 

\begin{figure}[htb]
\begin{center}
\includegraphics[scale=0.8]{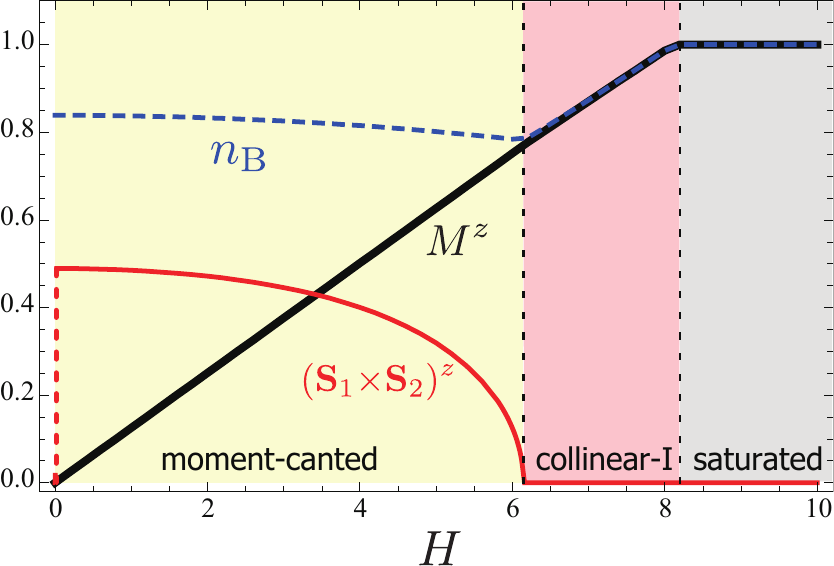}
\caption{(color online) Magnetization process exhibiting the anticipated canted 
configuration: $J_{\text{bl}}=2.0$, $J_{\text{bq}}=0.0$, $(t+u)/2=1.2$, 
$(t-u)/2=0.01$, 
$V_{\text{c}}=0.6$, $\mu=-0.5$.  At around $H=6.2$, spin state changes from 
the canted state to the collinear-I (with a kink in $n_{\text{B}}$).  
The vector chirality $(\bolS_{1}{\times}\bolS_{2})^{z}$ is finite in small-field region 
where the local magnetization assumes canted configuration (`moment-canted'). 
\label{fig:VC-vs-MH}}
\end{center}
\end{figure}

To seek for the possibility of the new phase with uniform chirality, 
we explored the enlarged parameter space and extensively carried out the variational analysis.  
It turned out that it is not easy 
to realize the canted state  (Fig.~\ref{fig:canted-AB}) at intermediate 
fillings (i.e., $0<n_{\text{B}}<1$); it is quite often masked by the above three 
collinear states or the 1/2-plateau.   However, careful choices of parameters 
can stabilize the canted state.  For instance, we show the magnetization process 
of the generalized ladder with $J_{\text{bl}}=2.0$, $J_{\text{bq}}=0.0$, 
$J_{\mathrm{A}}=(t+u)/2=1.3$, $J_{\mathrm{B}}=(t-u)/2=0.01$, 
$V_{\text{c}}=0.6$, $\mu=-0.5$ in 
Fig.~\ref{fig:VC-vs-MH}.   One can clearly see that uniform $z$-component 
of vector chirality is induced in the low-field region.  As has been mentioned above, 
the appearance of uniform vector chirality in the low-field phase implies  
the breaking of the chain-exchange $\mathbb{Z}_{2}$-symmetry. 

\subsection{Numerical results}
]
In order to investigate the possible appearance of vector chirality,
we now turn to numerical investigation using the DMRG
algorithm. \cite{White-92} Since the pure ring-exchange model does not
exhibit such a symmetry breaking, \cite{Hikihara-Y-08} and because the most
general spin ladder Hamiltonian contains too many parameters, we have
used the previous variational analysis as a guide.

\begin{figure}[h]
\begin{center}
\includegraphics[scale=0.3,clip]{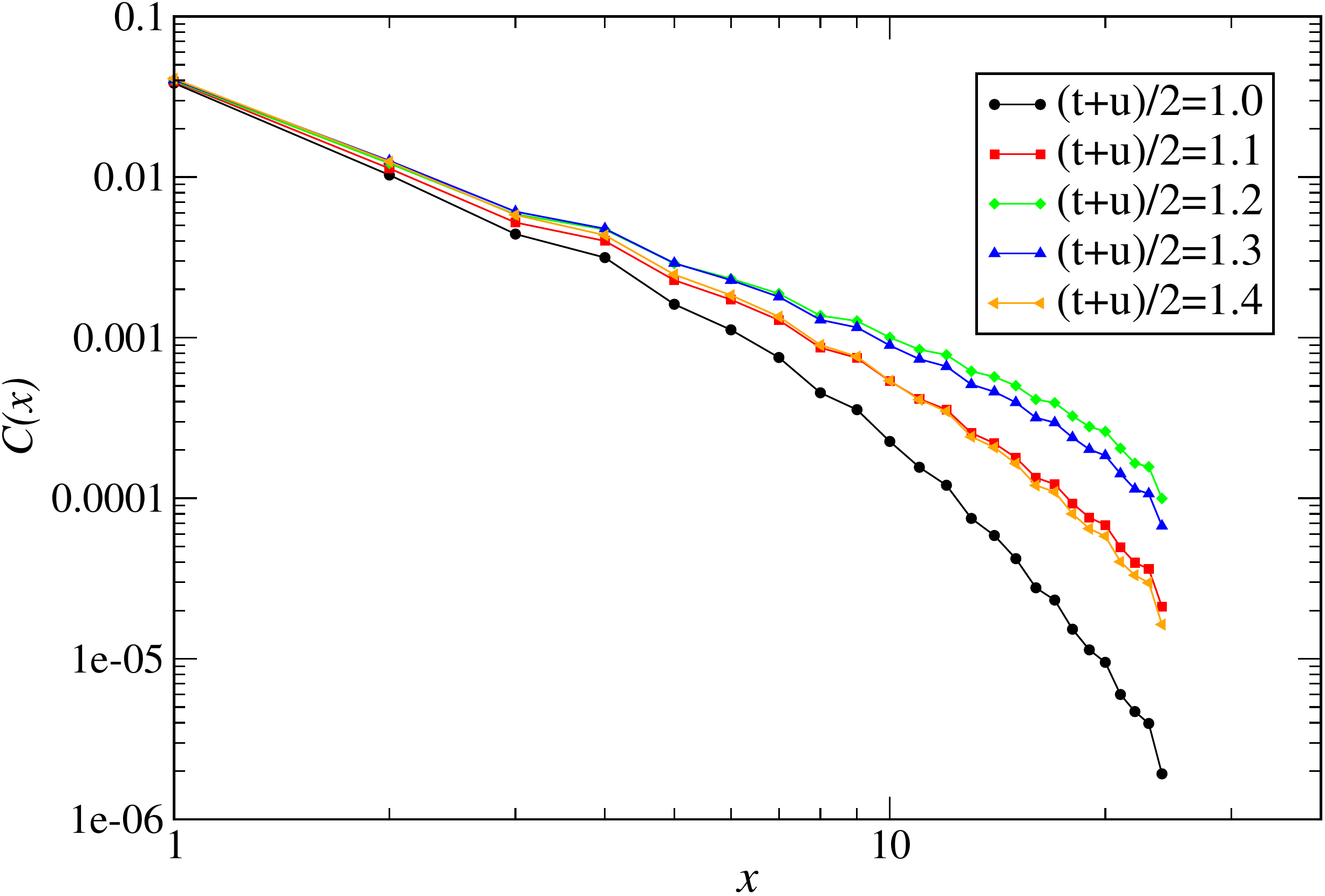}
\caption{(color online) 
Vector chirality correlations obtained on $2\times 32$ ladder with parameters given in the text. Magnetization was fixed at 1/4 of its saturation value. 
}
\label{fig:dmrg}
\end{center}
\end{figure}

Using the following parameters $J_{\text{bl}}=2.0$,
$J_{\text{bq}}=0.0$, $(t-u)/2=0.01$, $V_{\text{c}}=0.6$, and
$\mu=-0.5$, vector chirality was found 
to occur for $(t+u)/2=1.2$ in the variational approach (see Fig.~\ref{fig:VC-vs-MH}). 
In Fig.~\ref{fig:dmrg}, we plot the vector chirality correlation
\begin{equation}
C(x) = \langle (\bolS_{1,r}{\times}\bolS_{2,r})^z (\bolS_{1,r+x}{\times}\bolS_{2,r+x})^z \rangle
\end{equation}
obtained numerically with DMRG. We have studied $2\times L$ ladders with general
four-spin exchange (see Eq.~(\ref{eqn:general-ladder})) corresponding to
the above bosonic parameters, and we use open boundary conditions.
Correlations are averaged with $L/4 \leq r \leq 3L/4$. We keep up to
1200 states which is sufficient to get a discarded weight smaller than
$10^{-12}$.

As can be read from the plot, vector chirality correlations do exhibit a slight increase when $(t+u)/2 \sim 1.2$, corresponding to the optimal parameter found in the variational analysis, but this only indicates quasi long-range order, i.e. an algebraic decay. 

We have also tried to vary some of these parameters, or look at other magnetization values, but we have been unable so far to detect any evidence of long-range order. 

\section{Concluding Remarks}
\label{sec:concluding-remark}
In the present paper, we have investigated the phase diagram of the two-leg spin ladder with
a ring-exchange interaction. In this respect, we have developed a unifying approach 
which treats AF fluctuations and VC (or $p$-nematic) ones on the same footing.
This approach is based on the description in terms of a spin-1 hardcore boson 
operators that create the triplet states out of the singlet on each rung.  
The spin-chirality transformation, originally defined 
on the lattice spin operators,\cite{Hikihara-M-H-03,Momoi-H-N-H-03} has 
a simple physical meaning here through
the standard U(1) global gauge symmetry of the bosons.

Using a simple mean-field calculation on the boson operators,
we have mapped out the zero-temperature phase diagram of the two-leg spin ladder with
a ring-exchange interaction. 
This bosonic approach enables one to take into account quantum fluctuations in a semiclassical
fashion. In particular, we reproduce the main results of previous intensive numerical studies \cite{Lauchli-S-T-03,Hikihara-M-H-03}
and a low-energy description of the chirality-dominant phase is derived within 
the semiclassical analysis. 

The effect of a magnetic field can also be naturally analyzed in our bosonic approach.
We reveal the mutual induction phenomenon between spin and chirality degrees of freedom
with the emergence of a vector-chirality phase, e.g. 
with long-range ordering $(\bolS_{1}{\times}\bolS_{2})^{z} \ne 0$, 
by the application of a strong magnetic field along the $z$-axis.
Within a mean-field approach, we  found that the two-leg spin ladder with
a ring-exchange interaction (Eq. (\ref{eqn:ladder})) 
in a magnetic field does not support this exotic phase in full 
agreement with the DMRG study of Ref. \onlinecite{Hikihara-Y-08}.
In this respect, we have given conditions to observe this phase
by considering a  two-leg spin ladder with general four-spin exchange interactions.
Unfortunately, our preliminary DMRG results could not find any evidence of chirality long-range ordering, although a mean-field analysis predicts a finite window where uniform vector 
chirality along the magnetic field is stabilized.
Clearly extensive large-scale DMRG calculations are called for to fully investigate the six-parameters space of the problem and look for the  existence of 
the magnetic-field induced vector-chirality phase.

Finally, we briefly comment on a perspective.  
As we have mentioned above, our bosonic approach can be readily generalized 
to higher-dimensional systems made of two-spin clusters (spin dimers) 
like $S=1/2$ bilayer systems.  
However, the range of its applicability is much broader.   
In fact, as we will show elsewhere,\cite{Totsuka-C-L-unpub-12} 
even in the two-dimensional cases {\em without} such special structures, 
one can construct effective bosonic degrees of
freedom which encode the competition between spin and chirality degrees of freedom
in close parallel to the 1D case.  
Then, we can apply a semiclassical field-theory approach similar to what is described in 
this paper to such higher-dimensional systems as the spin-1/2 Heisenberg model 
on the square lattice with a ring-exchange interaction to capture the global structure of 
the phase diagram. 


\section*{Acknowledgements}
The authors thank F.~Essler, T.~Hikihara, A.~Kolezhuk, 
A.~L\"{a}uchli, C.~Lhuillier, F.~Mila, M.~Oshikawa, and M.~Sato for helpful discussions. 
We also wish to thank the organizers 
of the workshop/symposium 
{\it Yukawa International Seminar (YKIS) 2007} and 
{\it Topological Aspects of Solid State Physics} 
held at Yukawa Institute for Theoretical Physics 
where parts of this work were carried out. 
One of the authors (KT) is grateful 
to Max-Planck Institute for the physics of complex systems, 
where this work was completed, for the hospitality.   
The author (KT) was supported in part by 
Grant-in-Aids for Scientific Research 
(C) 18540372, (C) 20540375, and Priority Areas ``Novel States of Matter 
Induced by Frustration'' (No.19052003) from MEXT, Japan and 
by the global COE (GCOE) program `The next generation of 
physics, spun from universality and emergence' of Kyoto University. 
One of the authors (SC) would like to thank Institut Universitaire de France 
for financial support. 
\appendix
\section{Dimer coherent state path integral} 
\label{sec:path-integral}
In this Appendix, we present some technical details on the derivation 
of the low-energy effective actions for the RS and VC phases of the two-leg spin ladder
with a ring-exchange (\ref{eqn:ladder}).
\subsection{Berry phase term}
\label{sec:Berry-phase}
The quantum dynamics is generated by the Berry phase term which 
can be derived from the following overlap:
\begin{equation}
\langle s(t+\delta t), \bolt(t+\delta t)|s(t),\bolt(t)\rangle 
\approx 1+
\left\{
s\, \partial_{t}s^{\ast} + \bolt{\cdot}\partial_{t}\bolt^{\ast}
\right\} \delta t \; ,
\end{equation}
where $t$ and $t+\delta t$ denote two infinitesimally separated 
times. 
Plugging (\ref{eqn:gauge-fixed}) into the above, we obtain
\begin{multline}
\langle s(t+\delta t), \bolt(t+\delta t)|s(t),\bolt(t)\rangle \\
= 1+
i(\bolA{\cdot}\partial_{t}\bolB-\bolB{\cdot}\partial_{t}\bolA)\, 
\delta t \\
= 1 -\frac{1}{2}(\bolt^{\ast}{\cdot}\partial_{t}\bolt
- \bolt{\cdot}\partial_{t}\bolt^{\ast}) \delta t \; ,
\end{multline}
from which we can read off the Berry phase contribution to the 
(single-site) action:
\begin{equation}
{\cal S}_{\text{Berry}} 
= \int\! dt \frac{i}{2} 
(\bolt^{\ast}{\cdot}\partial_{t}\bolt
- \bolt{\cdot}\partial_{t}\bolt^{\ast}) 
= \int\!dt (\bolA{\cdot}\partial_{t}\bolB-\bolB{\cdot}\partial_{t}\bolA)
 \; .
\end{equation}
This Berry phase term is the first contribution of the general action (\ref{action}).
As is easily seen by rescaling: $\bolt \mapsto \sqrt{\rho}\,\bolt$ 
($|\bolt|=1$), 
this term generates the dynamics in the {\em transverse} direction, 
i.e. there is no time-derivative for the $\rho$-field which describe 
the fluctuations in the {\em longitudinal} direction.    

\subsection{Continuum limit}
\label{sec:App-continuum-limit}
On the basis of the fluctuation analysis in Sec. \ref{sec:low-energy-fluc}, 
we can derive a low-energy effective actions for the NAF-dominant- and 
the chirality-dominant phase.  
Since the $\bolA$-field develops in the former phase, we may make 
the following ansatz Eq.~(\ref{eqn:AB-by-Omega})
to parametrize the $\bolA$ and $\bolB$ fields:\begin{equation}
\begin{split}
&\bolA_{r}+i\, \bolB_{r} = 
\sqrt{\rho_{0}^{2}+a\, l_{\phi}(x)} \, 
(-1)^{r}\be^{i\phi(x)} \\
& \quad \times 
\left\{\sqrt{1-a^{2} \mathbf{l}^{2}_{\Omega}(x)}\, \bolOmega(x) + 
i a(\bolOmega{\times}\mathbf{l}_{\Omega}) \right\}  \; .
\end{split}  
\label{eqn:AB-by-omega-2}
\end{equation}
Clearly, $\bolOmega$ and $\phi$ respectively describe the transverse- and the phase 
fluctuations of the spin-1 boson.  
To guarantee the normalization condition $\VEV{n_{\text{B}}}=\rho_{0}^{2}$, 
we impose the following constraints:
\begin{equation}
|\bolOmega|=1 \; , \;\; \bolOmega \cdot \mathbf{l}_{\Omega}=0 \; .
\end{equation}  
The idea underlying the decomposition (\ref{eqn:AB-by-omega-2}) may be seen 
as follows.  First we note that Eq. (\ref{eqn:AB-by-omega-2}) is a close parallel of the expression 
(\ref{eqn:boson-by-AB}). Suppose that the vector field $\bolA$ has a finite length 
$\rho_{0}^{2}$ (as in NAF-dominant phase) and that it plays a role of the order parameter 
i.e. $\bolOmega \sim \bolA$.  The imaginary part $\bolB$ may be 
obtained by noting that the SU(2)-generator is given by 
\begin{equation}
\boldsymbol{L}_{r} = -2 \bolA_{r}{\times}\bolB_{r}
\end{equation}
(see Eq. (\ref{eqn:local-moment-by-AB})).   
As is suggested by counting the number of independent degrees of freedom, 
once the overall phase $\phi$ is singled out, 
the angle between $\bolA_r$ and $\bolB_r$ cannot be arbitrary; 
a convenient choice would be to take: $\bolA_r{\cdot} \bolB_r=0$.   
When $\bolA_r \sim \bolOmega_r$, $\bolB_r$ is expressed simply as:
\begin{equation}
\bolOmega_{r}{\times}\boldsymbol{L}_{r} 
= -2 \left\{
(\bolOmega_{r}{\cdot}\bolB_{r})\bolOmega_{r} - (\bolOmega_{r})^{2}\bolB_{r}
\right\} = 2\bolB_{r} \; (\perp \bolA_{r}) \; .
\end{equation}
With the identification: $\mathbf{l}_{\Omega}\sim \boldsymbol{L}/2$, one can see 
that the vectorial part of (\ref{eqn:AB-by-omega-2}) correctly reproduces 
the transformation properties of $\bolA+i \bolB$.  
The phase part $\sqrt{\rho_{0}^{2}+a\, l_{\phi}(x)} \, \be^{i\phi(x)}$ 
is easily guessed from the well-known result in the single-component Bose liquid 
\cite{Haldane-PRL-81}.  

Now let us plug the ansatz (\ref{eqn:AB-by-omega-2}) into 
the action (\ref{action}) and keep terms up to 
second order in the small fields $\mathbf{l}_{\Omega}$ and $l_{\phi}$.  
The constant part (\ref{eqn:eff-Ham-constant}) 
\begin{equation}
E_{0} = \rho_0^2 \left\{ \rho_0^2 (J+6 K_4 ) {-} 6 K_4\right\}
- \rho_0^2 (1-\rho_0^2)  (J-2 K_4) \cos (2 \phi_{0})  
\end{equation}
is minimized to give $\rho_{0}$ and $\phi_{0}$:
\begin{equation}
\begin{split}
& (\rho_{0},\phi_{0}) \\
& = 
\begin{cases}
\left( \frac{1}{2}\sqrt{\frac{J+4K_{4}}{J+2K_{4}}}  , \; 0 (\text{or }\pi) \right)\quad 
& \text{for } -1/4<K_{4}/J < 1/2 \\
\left( \frac{1}{4}\sqrt{\frac{-J+8K_{4}}{K_{4}}} , \; \frac{\pi}{2} (\text{or }\frac{3\pi}{2}) \right)
& \text{for } K_{4}/J > 1/2 \; .
\end{cases}
\end{split}
\end{equation}
Dropping the alternating sums and 
changing the sum over rungs to integrals
\begin{equation}
\sum_{r} \mapsto \frac{1}{a}\int\!dx  \; ,
\end{equation}
we obtain 
the following results:\footnote{After neglecting the longitudinal 
fluctuations, the charge part ${\cal H}_{\text{charge}}$ (\ref{eqn:classical-AB-charge}) merely 
contributes a constant.}
\begin{subequations}
\begin{equation}
{\cal S}_{\text{Berry}} =  \int\!dx\,dt \left\{
l_{\phi} (\partial_{t}\phi)
+ 2 \rho_0^2\, \mathbf{l}_{\Omega}{\cdot} [\bolOmega{\times}(\partial_{t}\bolOmega) ] 
 +\frac{\rho_0^2}{a} (\partial_{t}\phi)  \right\},
\end{equation}
\begin{equation}
\begin{split}
& {\cal S}_{\text{magnetic}} + {\cal S}_{\text{charge}} \\
& \quad \approx  - \left(2J +4K_4\right) \rho_0^4 a \int\!dx\,dt \,
 \left(\mathbf{l}_{\Omega}\right)^{2} -4K_{4}a \int\!dx\,dt \, l_{\phi}^{2} ,
 \end{split}
\end{equation}
and
\begin{equation}
\begin{split}
& {\cal S}_{\text{hopping}} \\ 
& \approx  
-\left\{ \left( \frac{J}{2}+K_4\right)\pm \left(\frac{J}{2}-K_4\right) \right\}a \\
& \quad \times  \rho_0^2 \left(1-\rho_0^2\right) \int\!dx\,dt \,\left\{   
\left[ \left(\partial_{x} \bolOmega\right)^{2}+(\partial_{x}\phi)^2    \right]
+2 l_{\phi}^{2}\right\} \\
& \quad  \mp 4a \rho_0^2\left(1-\rho_0^2\right) \left(\frac{J}{2}-K_4\right)   
  \int\!dx\,dt \, \left( \mathbf{l}_{\Omega}\right)^{2}  \; ,
\end{split}
\end{equation}
\end{subequations}
where the upper (lower) sign is chosen when $J>2K_{4}$ ($J<2K_{4}$). 
The Gaussian integration in $\mathbf{l}_{\Omega}$ and $l_{\phi}$ yields 
the following result:
\begin{equation}
\begin{split}
{\cal S}_{\text{eff}}[\bolOmega] = &
\frac{1}{2}\chi_{\text{s}} \int\!dx dt \,
(\partial_{t}\bolOmega)^{2} 
- \frac{1}{2}\rho_{\text{s}}
\int\!dx dt \, (\partial_{x} \bolOmega)^{2} \\
& +\frac{1}{2}\chi_{\text{c}} \int\!dx dt \,
(\partial_{t}\phi)^{2} 
- \frac{1}{2}\rho_{\text{c}}
\int\!dx dt \, (\partial_{x} \phi)^{2} \\
& + g(J,K_{4})  \int\!dx dt \, \cos(2\phi) \; .
\end{split}
\label{eqn:App-eff-action}
\end{equation}

When $J>2K_{4}$, the effective action ${\cal S}_{\text{eff}}[\bolOmega]$ describes 
the $(\pi,\pi)$ AF fluctuations and the phase fluctuations in 
the spin-chirality space. 
The five coupling constants $\chi_{\text{s,c}}$, $\rho_{\text{s,c}}$ 
and $g(J,K_{4})$ are given by:
\begin{subequations}
\begin{align}
& \chi_{\text{s}}=\frac{J+4K_{4}}{4a J(J+K_{4})} \; , \; 
\rho_{\text{s}}=\frac{J(J+4K_{4})(3J+4K_{4})a}{8(J+2K_{4})^{2}}, \\
& \chi_{\text{c}}=\frac{1}{4 a (J+2 K_4)} \; , \; 
\rho_{\text{c}}= \frac{J (J+4 K_4) (3 J+4 K_4)a}{8 (J+2 K_4)^2} ,\\
& g(J,K_4) = \frac{(J{-}2 K_4)(J{+}4K_4) (3 J{+}4 K_4)}{16 (J{+}2 K_4)^2a}  \;
(>0) .
\end{align} 
\label{eqn:chi-rho-A}
\end{subequations}   
Note that precisely at the self-dual point $J=2K_{4}$, $g(J,K_{4})=0$ 
and the spin-chirality fluctuations becomes gapless (i.e. Tomonaga-Luttinger like). 

In the chirality-dominant phase $K_{4}>J/2$ (or $\theta_{\text{sd}}<\theta<\pi/2$), 
$\phi$ is locked at $\phi=\pi/2,3\pi/2$ and 
$\bolOmega \sim (\bolS_{1}{\times}\bolS_{2})_{k=\pi}$.  
This short-range vector-chiral fluctuations at $k=\pi$ are governed 
by the non-linear sigma model (\ref{eqn:App-eff-action}) with 
\begin{subequations}
\begin{align}
& \chi_{\text{s}}=\frac{8K_{4}-J}{2a(4K_{4}{-}J)(4K_{4}{+}J)} \; , \;
\rho_{\text{s}}= \frac{(8K_{4}{-}J)(8K_{4}{+}J)a}{64 K_{4}} , \\
& \chi_{\text{c}}=\frac{1}{16 a K_4}  \; , \;
\rho_{\text{c}}= \frac{(8 K_4-J) (8 K_4+J)a}{64 K_4},  \\
& g(J,K_{4})= -\frac{(2 K_4{-}J)(8 K_4{-}J)(8 K_4{+}J)}{256 K_4^2 a} \; (<0)
\; .
\end{align}
\label{eqn:chi-rho-B}
\end{subequations}

In deriving the effective model for the F-nematic phase 
realized for $J+2K_{4}<0$,  
we use another parametrization (\ref{eqn:eqn:ansatz-F-nem}) which takes into account 
the transverse soft modes at $k=0$ and the staggered U(1)-symmetry 
for $J+2K_{4}=0$:
\begin{equation}
\begin{split}
&\bolA_{r}+i\, \bolB_{r} = 
\sqrt{\rho_{0}^{2}+a\, l_{\phi}(x)} \, 
\be^{(-1)^{r}i\phi(x)} \\
& \quad \times 
\left\{\sqrt{1-a^{2} \mathbf{l}^{2}_{\Omega}(x)}\, \bolOmega(x) + 
i a(\bolOmega{\times}\mathbf{l}_{\Omega}) \right\}  \; ,
\end{split}
\label{eqn:eqn:ansatz-F-nem-2}  
\end{equation}

where the two phenomenological parameters are now given by:
\begin{subequations}
\begin{align}
\begin{split}
& \chi_{\mathrm{s}} = \frac{3 J-4 K_4}{2 a \left(J^2+4 K_4 J-8 K_4^2\right)} \; , \\
& \rho_{\mathrm{s}} = 
-\frac{J \left(J-4 K_4\right) \left(3 J-4 K_4\right)a}{8 \left(J-2 K_4\right)^2}  \; .
\end{split}
\\
& \chi_{\text{c}} = \frac{1}{4a (2K_4 -J)} \; , \;\;
\rho_{\text{c}} = \frac{a (-J)(4K_4-J) (4K_4-3 J)}{8 (J-2K_4)^2} ,\\
& g(J,K_{4}) = -\frac{(J{+}2K_4)(4K_4{-}J) (4K_4{-}3J)}{16 a (J-2K_4)^2} \; (>0) .
\end{align}
\label{eqn:chi-rho-F-NEM}
\end{subequations}


\begin{thebibliography}{70}%
\makeatletter
\providecommand \@ifxundefined [1]{%
 \@ifx{#1\undefined}
}%
\providecommand \@ifnum [1]{%
 \ifnum #1\expandafter \@firstoftwo
 \else \expandafter \@secondoftwo
 \fi
}%
\providecommand \@ifx [1]{%
 \ifx #1\expandafter \@firstoftwo
 \else \expandafter \@secondoftwo
 \fi
}%
\providecommand \natexlab [1]{#1}%
\providecommand \enquote  [1]{``#1''}%
\providecommand \bibnamefont  [1]{#1}%
\providecommand \bibfnamefont [1]{#1}%
\providecommand \citenamefont [1]{#1}%
\providecommand \href@noop [0]{\@secondoftwo}%
\providecommand \href [0]{\begingroup \@sanitize@url \@href}%
\providecommand \@href[1]{\@@startlink{#1}\@@href}%
\providecommand \@@href[1]{\endgroup#1\@@endlink}%
\providecommand \@sanitize@url [0]{\catcode `\\12\catcode `\$12\catcode
  `\&12\catcode `\#12\catcode `\^12\catcode `\_12\catcode `\%12\relax}%
\providecommand \@@startlink[1]{}%
\providecommand \@@endlink[0]{}%
\providecommand \url  [0]{\begingroup\@sanitize@url \@url }%
\providecommand \@url [1]{\endgroup\@href {#1}{\urlprefix }}%
\providecommand \urlprefix  [0]{URL }%
\providecommand \Eprint [0]{\href }%
\providecommand \doibase [0]{http://dx.doi.org/}%
\providecommand \selectlanguage [0]{\@gobble}%
\providecommand \bibinfo  [0]{\@secondoftwo}%
\providecommand \bibfield  [0]{\@secondoftwo}%
\providecommand \translation [1]{[#1]}%
\providecommand \BibitemOpen [0]{}%
\providecommand \bibitemStop [0]{}%
\providecommand \bibitemNoStop [0]{.\EOS\space}%
\providecommand \EOS [0]{\spacefactor3000\relax}%
\providecommand \BibitemShut  [1]{\csname bibitem#1\endcsname}%
\let\auto@bib@innerbib\@empty
\bibitem [{\citenamefont {Thouless}(1965)}]{Thouless-65}%
  \BibitemOpen
  \bibfield  {author} {\bibinfo {author} {\bibfnamefont {D.~J.}\ \bibnamefont
  {Thouless}},\ }\href {http://stacks.iop.org/0370-1328/86/i=5/a=301}
  {\bibfield  {journal} {\bibinfo  {journal} {Proc. Phys.
  Soc. London}\ }\textbf {\bibinfo {volume} {86}},\ \bibinfo {pages}
  {893} (\bibinfo {year} {1965})}\BibitemShut {NoStop}%
\bibitem [{\citenamefont {Takahashi}(1977)}]{Takahashi-77}%
  \BibitemOpen
  \bibfield  {author} {\bibinfo {author} {\bibfnamefont {M.}~\bibnamefont
  {Takahashi}},\ }\href {http://stacks.iop.org/0022-3719/10/i=8/a=031}
  {\bibfield  {journal} {\bibinfo  {journal} {J. Phys. C: Solid State Physics}\
  }\textbf {\bibinfo {volume} {10}},\ \bibinfo {pages} {1289} (\bibinfo {year}
  {1977})}\BibitemShut {NoStop}%
\bibitem [{\citenamefont {MacDonald}\ \emph {et~al.}(1988)\citenamefont
  {MacDonald}, \citenamefont {Girvin},\ and\ \citenamefont
  {Yoshioka}}]{MacDonald-G-Y-88}%
  \BibitemOpen
  \bibfield  {author} {\bibinfo {author} {\bibfnamefont {A.}~\bibnamefont
  {MacDonald}}, \bibinfo {author} {\bibfnamefont {S.}~\bibnamefont {Girvin}}, \
  and\ \bibinfo {author} {\bibfnamefont {D.}~\bibnamefont {Yoshioka}},\ }\href
  {http://prb.aps.org/abstract/PRB/v37/i16/p9753_1} {\bibfield  {journal}
  {\bibinfo  {journal} {Phys. Rev. B}\ }\textbf {\bibinfo {volume} {37}},\
  \bibinfo {pages} {9753} (\bibinfo {year} {1988})}\BibitemShut {NoStop}%
\bibitem [{\citenamefont {Roger}\ \emph {et~al.}(1983)\citenamefont {Roger},
  \citenamefont {Hetherington},\ and\ \citenamefont {Delrieu}}]{Roger-H-D-83}%
  \BibitemOpen
  \bibfield  {author} {\bibinfo {author} {\bibfnamefont {M.}~\bibnamefont
  {Roger}}, \bibinfo {author} {\bibfnamefont {J.~H.}\ \bibnamefont
  {Hetherington}}, \ and\ \bibinfo {author} {\bibfnamefont {J.~M.}\
  \bibnamefont {Delrieu}},\ }\href
  {http://link.aps.org/doi/10.1103/RevModPhys.55.1} {\bibfield  {journal}
  {\bibinfo  {journal} {Rev. Mod. Phys.}\ }\textbf {\bibinfo {volume} {55}},\
  \bibinfo {pages} {1} (\bibinfo {year} {1983})}\BibitemShut {NoStop}%
\bibitem [{\citenamefont {Fukuyama}(2008)}]{Fukuyama-08}%
  \BibitemOpen
  \bibfield  {author} {\bibinfo {author} {\bibfnamefont {H.}~\bibnamefont
  {Fukuyama}},\ }\href {\doibase 10.1143/JPSJ.77.111013} {\bibfield  {journal}
  {\bibinfo  {journal} {J. Phys. Soc. Jpn.}\ }\textbf {\bibinfo {volume}
  {77}},\ \bibinfo {pages} {111013} (\bibinfo {year} {2008})}\BibitemShut
  {NoStop}%
\bibitem [{\citenamefont {Chakravarty}\ \emph {et~al.}(1999)\citenamefont
  {Chakravarty}, \citenamefont {Kivelson}, \citenamefont {Nayak},\ and\
  \citenamefont {Voelker}}]{Chakravarty-K-N-V-99}%
  \BibitemOpen
  \bibfield  {author} {\bibinfo {author} {\bibfnamefont {S.}~\bibnamefont
  {Chakravarty}}, \bibinfo {author} {\bibfnamefont {S.}~\bibnamefont
  {Kivelson}}, \bibinfo {author} {\bibfnamefont {C.}~\bibnamefont {Nayak}}, \
  and\ \bibinfo {author} {\bibfnamefont {K.}~\bibnamefont {Voelker}},\ }\href
  {http://www.tandfonline.com/doi/abs/10.1080/13642819908214845} {\bibfield
  {journal} {\bibinfo  {journal} {Philosophical Magazine Part B}\ }\textbf
  {\bibinfo {volume} {79}},\ \bibinfo {pages} {859} (\bibinfo {year}
  {1999})}\BibitemShut {NoStop}%
\bibitem [{\citenamefont {Shimizu}\ \emph {et~al.}(2003)\citenamefont
  {Shimizu}, \citenamefont {Miyagawa}, \citenamefont {Kanoda}, \citenamefont
  {Maesato},\ and\ \citenamefont {Saito}}]{Shimizu-M-K-M-S-03}%
  \BibitemOpen
  \bibfield  {author} {\bibinfo {author} {\bibfnamefont {Y.}~\bibnamefont
  {Shimizu}}, \bibinfo {author} {\bibfnamefont {K.}~\bibnamefont {Miyagawa}},
  \bibinfo {author} {\bibfnamefont {K.}~\bibnamefont {Kanoda}}, \bibinfo
  {author} {\bibfnamefont {M.}~\bibnamefont {Maesato}}, \ and\ \bibinfo
  {author} {\bibfnamefont {G.}~\bibnamefont {Saito}},\ }\href
  {http://link.aps.org/abstract/PRL/v91/e107001} {\bibfield  {journal}
  {\bibinfo  {journal} {Phys. Rev. Lett.}\ }\textbf {\bibinfo {volume} {91}},\
  \bibinfo {pages} {107001} (\bibinfo {year} {2003})}\BibitemShut {NoStop}%
\bibitem [{\citenamefont {Yamashita}\ \emph {et~al.}(2010)\citenamefont
  {Yamashita}, \citenamefont {Nakata}, \citenamefont {Senshu}, \citenamefont
  {Nagata}, \citenamefont {Yamamoto}, \citenamefont {Kato}, \citenamefont
  {Shibauchi},\ and\ \citenamefont {Matsuda}}]{Yamashita-etal-10}%
  \BibitemOpen
  \bibfield  {author} {\bibinfo {author} {\bibfnamefont {M.}~\bibnamefont
  {Yamashita}}, \bibinfo {author} {\bibfnamefont {N.}~\bibnamefont {Nakata}},
  \bibinfo {author} {\bibfnamefont {Y.}~\bibnamefont {Senshu}}, \bibinfo
  {author} {\bibfnamefont {M.}~\bibnamefont {Nagata}}, \bibinfo {author}
  {\bibfnamefont {H.~M.}\ \bibnamefont {Yamamoto}}, \bibinfo {author}
  {\bibfnamefont {R.}~\bibnamefont {Kato}}, \bibinfo {author} {\bibfnamefont
  {T.}~\bibnamefont {Shibauchi}}, \ and\ \bibinfo {author} {\bibfnamefont
  {Y.}~\bibnamefont {Matsuda}},\ }\href {\doibase 10.1126/science.1188200}
  {\bibfield  {journal} {\bibinfo  {journal} {Science}\ }\textbf {\bibinfo
  {volume} {328}},\ \bibinfo {pages} {1246} (\bibinfo {year}
  {2010})}\BibitemShut {NoStop}%
\bibitem [{\citenamefont {Coldea}\ \emph {et~al.}(2001)\citenamefont {Coldea},
  \citenamefont {Hayden}, \citenamefont {Aeppli}, \citenamefont {Perring},
  \citenamefont {Frost}, \citenamefont {Mason}, \citenamefont {Cheong},\ and\
  \citenamefont {Fisk}}]{Coldea-H-A-P-F-M-C-F-01}%
  \BibitemOpen
  \bibfield  {author} {\bibinfo {author} {\bibfnamefont {R.}~\bibnamefont
  {Coldea}}, \bibinfo {author} {\bibfnamefont {S.~M.}\ \bibnamefont {Hayden}},
  \bibinfo {author} {\bibfnamefont {G.}~\bibnamefont {Aeppli}}, \bibinfo
  {author} {\bibfnamefont {T.~G.}\ \bibnamefont {Perring}}, \bibinfo {author}
  {\bibfnamefont {C.~D.}\ \bibnamefont {Frost}}, \bibinfo {author}
  {\bibfnamefont {T.~E.}\ \bibnamefont {Mason}}, \bibinfo {author}
  {\bibfnamefont {S.-W.}\ \bibnamefont {Cheong}}, \ and\ \bibinfo {author}
  {\bibfnamefont {Z.}~\bibnamefont {Fisk}},\ }\href
  {http://link.aps.org/abstract/PRL/v86/p5377} {\bibfield  {journal} {\bibinfo
  {journal} {Phys. Rev. Lett.}\ }\textbf {\bibinfo {volume} {86}},\ \bibinfo
  {pages} {5377} (\bibinfo {year} {2001})}\BibitemShut {NoStop}%
\bibitem [{\citenamefont {Brehmer}\ \emph {et~al.}(1999)\citenamefont
  {Brehmer}, \citenamefont {Mikeska}, \citenamefont {Muller}, \citenamefont
  {Nagaosa},\ and\ \citenamefont {Uchida}}]{Brehmer-M-M-N-U-99}%
  \BibitemOpen
  \bibfield  {author} {\bibinfo {author} {\bibfnamefont {S.}~\bibnamefont
  {Brehmer}}, \bibinfo {author} {\bibfnamefont {H.-J.}\ \bibnamefont
  {Mikeska}}, \bibinfo {author} {\bibfnamefont {M.}~\bibnamefont {Muller}},
  \bibinfo {author} {\bibfnamefont {N.}~\bibnamefont {Nagaosa}}, \ and\
  \bibinfo {author} {\bibfnamefont {S.}~\bibnamefont {Uchida}},\ }\href
  {http://link.aps.org/abstract/PRB/v60/p329} {\bibfield  {journal} {\bibinfo
  {journal} {Phys. Rev. B}\ }\textbf {\bibinfo {volume} {60}},\ \bibinfo
  {pages} {329} (\bibinfo {year} {1999})}\BibitemShut {NoStop}%
\bibitem [{\citenamefont {Matsuda}\ \emph {et~al.}(2000)\citenamefont
  {Matsuda}, \citenamefont {Katsumata}, \citenamefont {Eccleston},
  \citenamefont {Brehmer},\ and\ \citenamefont {Mikeska}}]{Matsuda-K-E-B-M-00}%
  \BibitemOpen
  \bibfield  {author} {\bibinfo {author} {\bibfnamefont {M.}~\bibnamefont
  {Matsuda}}, \bibinfo {author} {\bibfnamefont {K.}~\bibnamefont {Katsumata}},
  \bibinfo {author} {\bibfnamefont {R.~S.}\ \bibnamefont {Eccleston}}, \bibinfo
  {author} {\bibfnamefont {S.}~\bibnamefont {Brehmer}}, \ and\ \bibinfo
  {author} {\bibfnamefont {H.-J.}\ \bibnamefont {Mikeska}},\ }\href
  {http://link.aps.org/abstract/PRB/v62/p8903} {\bibfield  {journal} {\bibinfo
  {journal} {Phys. Rev. B}\ }\textbf {\bibinfo {volume} {62}},\ \bibinfo
  {pages} {8903} (\bibinfo {year} {2000})}\BibitemShut {NoStop}%
\bibitem [{\citenamefont {Nunner}\ \emph {et~al.}(2002)\citenamefont {Nunner},
  \citenamefont {Brune}, \citenamefont {Kopp}, \citenamefont {Windt},\ and\
  \citenamefont {Gr\"{u}ninger}}]{Nunner-B-K-W-G-02}%
  \BibitemOpen
  \bibfield  {author} {\bibinfo {author} {\bibfnamefont {T.~S.}\ \bibnamefont
  {Nunner}}, \bibinfo {author} {\bibfnamefont {P.}~\bibnamefont {Brune}},
  \bibinfo {author} {\bibfnamefont {T.}~\bibnamefont {Kopp}}, \bibinfo {author}
  {\bibfnamefont {M.}~\bibnamefont {Windt}}, \ and\ \bibinfo {author}
  {\bibfnamefont {M.}~\bibnamefont {Gr\"{u}ninger}},\ }\href
  {http://link.aps.org/abstract/PRB/v66/e180404} {\bibfield  {journal}
  {\bibinfo  {journal} {Phys. Rev. B}\ }\textbf {\bibinfo {volume} {66}},\
  \bibinfo {pages} {180404} (\bibinfo {year} {2002})}\BibitemShut {NoStop}%
\bibitem [{\citenamefont {Calzado}\ \emph {et~al.}(2003)\citenamefont
  {Calzado}, \citenamefont {de~Graaf}, \citenamefont {Bordas}, \citenamefont
  {Caballol},\ and\ \citenamefont {Malrieu}}]{Calzado-G-B-C-M-03}%
  \BibitemOpen
  \bibfield  {author} {\bibinfo {author} {\bibfnamefont {C.~J.}\ \bibnamefont
  {Calzado}}, \bibinfo {author} {\bibfnamefont {C.}~\bibnamefont {de~Graaf}},
  \bibinfo {author} {\bibfnamefont {E.}~\bibnamefont {Bordas}}, \bibinfo
  {author} {\bibfnamefont {R.}~\bibnamefont {Caballol}}, \ and\ \bibinfo
  {author} {\bibfnamefont {J.-P.}\ \bibnamefont {Malrieu}},\ }\href
  {http://prb.aps.org/abstract/PRB/v67/i13/e132409} {\bibfield  {journal}
  {\bibinfo  {journal} {Phys. Rev. B}\ }\textbf {\bibinfo {volume} {67}},\
  \bibinfo {pages} {132409} (\bibinfo {year} {2003})}\BibitemShut {NoStop}%
\bibitem [{\citenamefont {Schmidt}\ and\ \citenamefont
  {Uhrig}(2005)}]{Schmidt-U-05}%
  \BibitemOpen
  \bibfield  {author} {\bibinfo {author} {\bibfnamefont {K.~P.}\ \bibnamefont
  {Schmidt}}\ and\ \bibinfo {author} {\bibfnamefont {G.~S.}\ \bibnamefont
  {Uhrig}},\ }\href
  {http://search.ebscohost.com/login.aspx?direct=true&db=a9h&AN=18796527&site=%
ehost-live} {\bibfield  {journal} {\bibinfo  {journal} {Mod. Phys. Lett. B}\
  }\textbf {\bibinfo {volume} {19}},\ \bibinfo {pages} {1179 } (\bibinfo {year}
  {2005})}\BibitemShut {NoStop}%
\bibitem [{\citenamefont {Notbohm}\ \emph {et~al.}(2007)\citenamefont
  {Notbohm}, \citenamefont {Ribeiro}, \citenamefont {Lake}, \citenamefont
  {Tennant}, \citenamefont {Schmidt}, \citenamefont {Uhrig}, \citenamefont
  {Hess}, \citenamefont {Klingeler}, \citenamefont {Behr}, \citenamefont
  {Buchner}, \citenamefont {Reehuis}, \citenamefont {Bewley}, \citenamefont
  {Frost}, \citenamefont {Manuel},\ and\ \citenamefont
  {Eccleston}}]{Notbohm-etal-07}%
  \BibitemOpen
  \bibfield  {author} {\bibinfo {author} {\bibfnamefont {S.}~\bibnamefont
  {Notbohm}}, \bibinfo {author} {\bibfnamefont {P.}~\bibnamefont {Ribeiro}},
  \bibinfo {author} {\bibfnamefont {B.}~\bibnamefont {Lake}}, \bibinfo {author}
  {\bibfnamefont {D.~A.}\ \bibnamefont {Tennant}}, \bibinfo {author}
  {\bibfnamefont {K.~P.}\ \bibnamefont {Schmidt}}, \bibinfo {author}
  {\bibfnamefont {G.~S.}\ \bibnamefont {Uhrig}}, \bibinfo {author}
  {\bibfnamefont {C.}~\bibnamefont {Hess}}, \bibinfo {author} {\bibfnamefont
  {R.}~\bibnamefont {Klingeler}}, \bibinfo {author} {\bibfnamefont
  {G.}~\bibnamefont {Behr}}, \bibinfo {author} {\bibfnamefont {B.}~\bibnamefont
  {Buchner}}, \bibinfo {author} {\bibfnamefont {M.}~\bibnamefont {Reehuis}},
  \bibinfo {author} {\bibfnamefont {R.~I.}\ \bibnamefont {Bewley}}, \bibinfo
  {author} {\bibfnamefont {C.~D.}\ \bibnamefont {Frost}}, \bibinfo {author}
  {\bibfnamefont {P.}~\bibnamefont {Manuel}}, \ and\ \bibinfo {author}
  {\bibfnamefont {R.~S.}\ \bibnamefont {Eccleston}},\ }\href
  {http://link.aps.org/abstract/PRL/v98/e027403} {\bibfield  {journal}
  {\bibinfo  {journal} {Phys. Rev. Lett.}\ }\textbf {\bibinfo {volume} {98}},\
  \bibinfo {pages} {027403} (\bibinfo {year} {2007})}\BibitemShut {NoStop}%
\bibitem [{\citenamefont {Bordas}\ \emph {et~al.}(2005)\citenamefont {Bordas},
  \citenamefont {de~Graaf}, \citenamefont {Caballol},\ and\ \citenamefont
  {Calzado}}]{Bordas-G-C-C-05}%
  \BibitemOpen
  \bibfield  {author} {\bibinfo {author} {\bibfnamefont {E.}~\bibnamefont
  {Bordas}}, \bibinfo {author} {\bibfnamefont {C.}~\bibnamefont {de~Graaf}},
  \bibinfo {author} {\bibfnamefont {R.}~\bibnamefont {Caballol}}, \ and\
  \bibinfo {author} {\bibfnamefont {C.~J.}\ \bibnamefont {Calzado}},\ }\href
  {http://link.aps.org/doi/10.1103/PhysRevB.71.045108} {\bibfield  {journal}
  {\bibinfo  {journal} {Phys. Rev. B}\ }\textbf {\bibinfo {volume} {71}},\
  \bibinfo {pages} {045108} (\bibinfo {year} {2005})}\BibitemShut {NoStop}%
\bibitem [{\citenamefont {Lake}\ \emph {et~al.}(2010)\citenamefont {Lake},
  \citenamefont {Tsvelik}, \citenamefont {Notbohm}, \citenamefont
  {Alan~Tennant}, \citenamefont {Perring}, \citenamefont {Reehuis},
  \citenamefont {Sekar}, \citenamefont {Krabbes},\ and\ \citenamefont
  {Buchner}}]{Lake-etal-10}%
  \BibitemOpen
  \bibfield  {author} {\bibinfo {author} {\bibfnamefont {B.}~\bibnamefont
  {Lake}}, \bibinfo {author} {\bibfnamefont {A.~M.}\ \bibnamefont {Tsvelik}},
  \bibinfo {author} {\bibfnamefont {S.}~\bibnamefont {Notbohm}}, \bibinfo
  {author} {\bibfnamefont {D.}~\bibnamefont {Alan~Tennant}}, \bibinfo {author}
  {\bibfnamefont {T.~G.}\ \bibnamefont {Perring}}, \bibinfo {author}
  {\bibfnamefont {M.}~\bibnamefont {Reehuis}}, \bibinfo {author} {\bibfnamefont
  {C.}~\bibnamefont {Sekar}}, \bibinfo {author} {\bibfnamefont
  {G.}~\bibnamefont {Krabbes}}, \ and\ \bibinfo {author} {\bibfnamefont
  {B.}~\bibnamefont {Buchner}},\ }\href {http://dx.doi.org/10.1038/nphys1462}
  {\bibfield  {journal} {\bibinfo  {journal} {Nat Phys}\ }\textbf {\bibinfo
  {volume} {6}},\ \bibinfo {pages} {50} (\bibinfo {year} {2010})}\BibitemShut
  {NoStop}%
\bibitem [{\citenamefont {Mizuno}\ \emph {et~al.}(1999)\citenamefont {Mizuno},
  \citenamefont {Tohyama},\ and\ \citenamefont {Maekawa}}]{Mizuno-T-M-99}%
  \BibitemOpen
  \bibfield  {author} {\bibinfo {author} {\bibfnamefont {Y.}~\bibnamefont
  {Mizuno}}, \bibinfo {author} {\bibfnamefont {T.}~\bibnamefont {Tohyama}}, \
  and\ \bibinfo {author} {\bibfnamefont {S.}~\bibnamefont {Maekawa}},\ }\href
  {http://dx.doi.org/10.1023/A:1022577918657} {\bibfield  {journal} {\bibinfo
  {journal} {J. Low Temp. Phys.}\ }\textbf {\bibinfo {volume} {117}},\ \bibinfo
  {pages} {389} (\bibinfo {year} {1999})},\ \bibinfo {note}
  {10.1023/A:1022577918657}\BibitemShut {NoStop}%
\bibitem [{\citenamefont {L\"{a}uchli}\ \emph {et~al.}(2005)\citenamefont
  {L\"{a}uchli}, \citenamefont {Domenge}, \citenamefont {Lhuillier},
  \citenamefont {Sindzingre},\ and\ \citenamefont
  {Troyer}}]{Lauchli-D-L-S-T-05}%
  \BibitemOpen
  \bibfield  {author} {\bibinfo {author} {\bibfnamefont {A.}~\bibnamefont
  {L\"{a}uchli}}, \bibinfo {author} {\bibfnamefont {J.~C.}\ \bibnamefont
  {Domenge}}, \bibinfo {author} {\bibfnamefont {C.}~\bibnamefont {Lhuillier}},
  \bibinfo {author} {\bibfnamefont {P.}~\bibnamefont {Sindzingre}}, \ and\
  \bibinfo {author} {\bibfnamefont {M.}~\bibnamefont {Troyer}},\ }\href
  {http://link.aps.org/abstract/PRL/v95/e137206} {\bibfield  {journal}
  {\bibinfo  {journal} {Phys. Rev. Lett.}\ }\textbf {\bibinfo {volume} {95}},\
  \bibinfo {pages} {137206} (\bibinfo {year} {2005})}\BibitemShut {NoStop}%
\bibitem [{\citenamefont {Shannon}\ \emph {et~al.}(2006)\citenamefont
  {Shannon}, \citenamefont {Momoi},\ and\ \citenamefont
  {Sindzingre}}]{Shannon-M-S-06}%
  \BibitemOpen
  \bibfield  {author} {\bibinfo {author} {\bibfnamefont {N.}~\bibnamefont
  {Shannon}}, \bibinfo {author} {\bibfnamefont {T.}~\bibnamefont {Momoi}}, \
  and\ \bibinfo {author} {\bibfnamefont {P.}~\bibnamefont {Sindzingre}},\
  }\href {http://link.aps.org/abstract/PRL/v96/e027213} {\bibfield  {journal}
  {\bibinfo  {journal} {Phys. Rev. Lett.}\ }\textbf {\bibinfo {volume} {96}},\
  \bibinfo {pages} {027213} (\bibinfo {year} {2006})}\BibitemShut {NoStop}%
\bibitem [{\citenamefont {Momoi}\ \emph {et~al.}(2006)\citenamefont {Momoi},
  \citenamefont {Sindzingre},\ and\ \citenamefont {Shannon}}]{Momoi-S-S-06}%
  \BibitemOpen
  \bibfield  {author} {\bibinfo {author} {\bibfnamefont {T.}~\bibnamefont
  {Momoi}}, \bibinfo {author} {\bibfnamefont {P.}~\bibnamefont {Sindzingre}}, \
  and\ \bibinfo {author} {\bibfnamefont {N.}~\bibnamefont {Shannon}},\ }\href
  {http://link.aps.org/doi/10.1103/PhysRevLett.97.257204} {\bibfield  {journal}
  {\bibinfo  {journal} {Phys. Rev. Lett.}\ }\textbf {\bibinfo {volume} {97}},\
  \bibinfo {pages} {257204} (\bibinfo {year} {2006})}\BibitemShut {NoStop}%
\bibitem [{\citenamefont {Senthil}\ \emph {et~al.}(2004)\citenamefont
  {Senthil}, \citenamefont {Vishwanath}, \citenamefont {Balents}, \citenamefont
  {Sachdev},\ and\ \citenamefont {Fisher}}]{Senthil-V-B-S-F-04}%
  \BibitemOpen
  \bibfield  {author} {\bibinfo {author} {\bibfnamefont {T.}~\bibnamefont
  {Senthil}}, \bibinfo {author} {\bibfnamefont {A.}~\bibnamefont {Vishwanath}},
  \bibinfo {author} {\bibfnamefont {L.}~\bibnamefont {Balents}}, \bibinfo
  {author} {\bibfnamefont {S.}~\bibnamefont {Sachdev}}, \ and\ \bibinfo
  {author} {\bibfnamefont {M.~P.~A.}\ \bibnamefont {Fisher}},\ }\href
  {http://www.sciencemag.org/cgi/content/abstract/303/5663/1490} {\bibfield
  {journal} {\bibinfo  {journal} {Science}\ }\textbf {\bibinfo {volume}
  {303}},\ \bibinfo {pages} {1490} (\bibinfo {year} {2004})}\BibitemShut
  {NoStop}%
\bibitem [{\citenamefont {T.~Senthil}\ and\ \citenamefont
  {Fisher}(2004)}]{Senthil-B-S-V-F-04}%
  \BibitemOpen
  \bibfield  {author} {\bibinfo {author} {\bibfnamefont {S.~S. A.~V.}\
  \bibnamefont {T.~Senthil}, \bibfnamefont {Leon~Balents}}\ and\ \bibinfo
  {author} {\bibfnamefont {M.}~\bibnamefont {Fisher}},\ }\href
  {http://link.aps.org/abstract/PRB/v70/e144407} {\bibfield  {journal}
  {\bibinfo  {journal} {Phys. Rev. B}\ }\textbf {\bibinfo {volume} {70}},\
  \bibinfo {pages} {144407} (\bibinfo {year} {2004})}\BibitemShut {NoStop}%
\bibitem [{\citenamefont {Sandvik}(2007)}]{Sandvik-07}%
  \BibitemOpen
  \bibfield  {author} {\bibinfo {author} {\bibfnamefont {A.~W.}\ \bibnamefont
  {Sandvik}},\ }\href {http://link.aps.org/abstract/PRL/v98/e227202} {\bibfield
   {journal} {\bibinfo  {journal} {Phys. Rev. Lett.}\ }\textbf {\bibinfo
  {volume} {98}},\ \bibinfo {pages} {227202} (\bibinfo {year}
  {2007})}\BibitemShut {NoStop}%
\bibitem [{\citenamefont {Sandvik}(2010)}]{Sandvik-10}%
  \BibitemOpen
  \bibfield  {author} {\bibinfo {author} {\bibfnamefont {A.~W.}\ \bibnamefont
  {Sandvik}},\ }\href {http://link.aps.org/doi/10.1103/PhysRevLett.104.177201}
  {\bibfield  {journal} {\bibinfo  {journal} {Phys. Rev. Lett.}\ }\textbf
  {\bibinfo {volume} {104}},\ \bibinfo {pages} {177201} (\bibinfo {year}
  {2010})}\BibitemShut {NoStop}%
\bibitem [{\citenamefont {Sheng}\ \emph {et~al.}(2009)\citenamefont {Sheng},
  \citenamefont {Motrunich},\ and\ \citenamefont {Fisher}}]{Sheng-M-F-09}%
  \BibitemOpen
  \bibfield  {author} {\bibinfo {author} {\bibfnamefont {D.~N.}\ \bibnamefont
  {Sheng}}, \bibinfo {author} {\bibfnamefont {O.~I.}\ \bibnamefont
  {Motrunich}}, \ and\ \bibinfo {author} {\bibfnamefont {M.~P.~A.}\
  \bibnamefont {Fisher}},\ }\href
  {http://link.aps.org/abstract/PRB/v79/e205112} {\bibfield  {journal}
  {\bibinfo  {journal} {Phys. Rev. B}\ }\textbf {\bibinfo {volume} {79}},\
  \bibinfo {pages} {205112} (\bibinfo {year} {2009})}\BibitemShut {NoStop}%
\bibitem [{\citenamefont {Block}\ \emph {et~al.}(2011)\citenamefont {Block},
  \citenamefont {Mishmash}, \citenamefont {Kaul}, \citenamefont {Sheng},
  \citenamefont {Motrunich},\ and\ \citenamefont
  {Fisher}}]{Block-M-K-S-M-F-11}%
  \BibitemOpen
  \bibfield  {author} {\bibinfo {author} {\bibfnamefont {M.~S.}\ \bibnamefont
  {Block}}, \bibinfo {author} {\bibfnamefont {R.~V.}\ \bibnamefont {Mishmash}},
  \bibinfo {author} {\bibfnamefont {R.~K.}\ \bibnamefont {Kaul}}, \bibinfo
  {author} {\bibfnamefont {D.~N.}\ \bibnamefont {Sheng}}, \bibinfo {author}
  {\bibfnamefont {O.~I.}\ \bibnamefont {Motrunich}}, \ and\ \bibinfo {author}
  {\bibfnamefont {M.~P.~A.}\ \bibnamefont {Fisher}},\ }\href
  {http://link.aps.org/doi/10.1103/PhysRevLett.106.046402} {\bibfield
  {journal} {\bibinfo  {journal} {Phys. Rev. Lett.}\ }\textbf {\bibinfo
  {volume} {106}},\ \bibinfo {pages} {046402} (\bibinfo {year}
  {2011})}\BibitemShut {NoStop}%
\bibitem [{\citenamefont {Misguich}\ \emph {et~al.}(1998)\citenamefont
  {Misguich}, \citenamefont {Bernu}, \citenamefont {Lhuillier},\ and\
  \citenamefont {Waldtmann}}]{Misguich-B-L-W-98}%
  \BibitemOpen
  \bibfield  {author} {\bibinfo {author} {\bibfnamefont {G.}~\bibnamefont
  {Misguich}}, \bibinfo {author} {\bibfnamefont {B.}~\bibnamefont {Bernu}},
  \bibinfo {author} {\bibfnamefont {C.}~\bibnamefont {Lhuillier}}, \ and\
  \bibinfo {author} {\bibfnamefont {C.}~\bibnamefont {Waldtmann}},\ }\href
  {http://link.aps.org/abstract/PRL/v81/p1098} {\bibfield  {journal} {\bibinfo
  {journal} {Phys. Rev. Lett.}\ }\textbf {\bibinfo {volume} {81}},\ \bibinfo
  {pages} {1098} (\bibinfo {year} {1998})}\BibitemShut {NoStop}%
\bibitem [{\citenamefont {Freedman}\ \emph {et~al.}(2005)\citenamefont
  {Freedman}, \citenamefont {Nayak},\ and\ \citenamefont
  {Shtengel}}]{Freedman-N-S-05}%
  \BibitemOpen
  \bibfield  {author} {\bibinfo {author} {\bibfnamefont {M.}~\bibnamefont
  {Freedman}}, \bibinfo {author} {\bibfnamefont {C.}~\bibnamefont {Nayak}}, \
  and\ \bibinfo {author} {\bibfnamefont {K.}~\bibnamefont {Shtengel}},\ }\href
  {http://link.aps.org/doi/10.1103/PhysRevLett.94.066401} {\bibfield  {journal}
  {\bibinfo  {journal} {Phys. Rev. Lett.}\ }\textbf {\bibinfo {volume} {94}},\
  \bibinfo {pages} {066401} (\bibinfo {year} {2005})}\BibitemShut {NoStop}%
\bibitem [{\citenamefont {M{\"{u}}ller}\ \emph {et~al.}(2002)\citenamefont
  {M{\"{u}}ller}, \citenamefont {Vekua},\ and\ \citenamefont
  {Mikeska}}]{Muller-V-M-02}%
  \BibitemOpen
  \bibfield  {author} {\bibinfo {author} {\bibfnamefont {M.}~\bibnamefont
  {M{\"{u}}ller}}, \bibinfo {author} {\bibfnamefont {T.}~\bibnamefont {Vekua}},
  \ and\ \bibinfo {author} {\bibfnamefont {H.-J.}\ \bibnamefont {Mikeska}},\
  }\href {http://link.aps.org/doi/10.1103/PhysRevB.66.134423} {\bibfield
  {journal} {\bibinfo  {journal} {Phys. Rev. B}\ }\textbf {\bibinfo {volume}
  {66}},\ \bibinfo {pages} {134423} (\bibinfo {year} {2002})}\BibitemShut
  {NoStop}%
\bibitem [{\citenamefont {L\"{a}uchli}\ \emph {et~al.}(2003)\citenamefont
  {L\"{a}uchli}, \citenamefont {Schmid},\ and\ \citenamefont
  {Troyer}}]{Lauchli-S-T-03}%
  \BibitemOpen
  \bibfield  {author} {\bibinfo {author} {\bibfnamefont {A.}~\bibnamefont
  {L\"{a}uchli}}, \bibinfo {author} {\bibfnamefont {G.}~\bibnamefont {Schmid}},
  \ and\ \bibinfo {author} {\bibfnamefont {M.}~\bibnamefont {Troyer}},\ }\href
  {http://link.aps.org/abstract/PRB/v67/e100409} {\bibfield  {journal}
  {\bibinfo  {journal} {Phys. Rev. B}\ }\textbf {\bibinfo {volume} {67}},\
  \bibinfo {pages} {100409} (\bibinfo {year} {2003})}\BibitemShut {NoStop}%
\bibitem [{\citenamefont {Hikihara}\ \emph {et~al.}(2003)\citenamefont
  {Hikihara}, \citenamefont {Momoi},\ and\ \citenamefont
  {Hu}}]{Hikihara-M-H-03}%
  \BibitemOpen
  \bibfield  {author} {\bibinfo {author} {\bibfnamefont {T.}~\bibnamefont
  {Hikihara}}, \bibinfo {author} {\bibfnamefont {T.}~\bibnamefont {Momoi}}, \
  and\ \bibinfo {author} {\bibfnamefont {X.}~\bibnamefont {Hu}},\ }\href
  {http://link.aps.org/abstract/PRL/v90/e087204} {\bibfield  {journal}
  {\bibinfo  {journal} {Phys. Rev. Lett.}\ }\textbf {\bibinfo {volume} {90}},\
  \bibinfo {pages} {087204} (\bibinfo {year} {2003})}\BibitemShut {NoStop}%
\bibitem [{\citenamefont {Momoi}\ \emph {et~al.}(2003)\citenamefont {Momoi},
  \citenamefont {Hikihara}, \citenamefont {Nakamura},\ and\ \citenamefont
  {Hu}}]{Momoi-H-N-H-03}%
  \BibitemOpen
  \bibfield  {author} {\bibinfo {author} {\bibfnamefont {T.}~\bibnamefont
  {Momoi}}, \bibinfo {author} {\bibfnamefont {T.}~\bibnamefont {Hikihara}},
  \bibinfo {author} {\bibfnamefont {M.}~\bibnamefont {Nakamura}}, \ and\
  \bibinfo {author} {\bibfnamefont {X.}~\bibnamefont {Hu}},\ }\href
  {http://link.aps.org/abstract/PRB/v67/e174410} {\bibfield  {journal}
  {\bibinfo  {journal} {Phys. Rev. B}\ }\textbf {\bibinfo {volume} {67}},\
  \bibinfo {pages} {174410} (\bibinfo {year} {2003})}\BibitemShut {NoStop}%
\bibitem [{\citenamefont {Gritsev}\ \emph {et~al.}(2004)\citenamefont
  {Gritsev}, \citenamefont {Normand},\ and\ \citenamefont
  {Baeriswyl}}]{Gritsev-N-B-04}%
  \BibitemOpen
  \bibfield  {author} {\bibinfo {author} {\bibfnamefont {V.}~\bibnamefont
  {Gritsev}}, \bibinfo {author} {\bibfnamefont {B.}~\bibnamefont {Normand}}, \
  and\ \bibinfo {author} {\bibfnamefont {D.}~\bibnamefont {Baeriswyl}},\ }\href
  {http://link.aps.org/doi/10.1103/PhysRevB.69.094431} {\bibfield  {journal}
  {\bibinfo  {journal} {Phys. Rev. B}\ }\textbf {\bibinfo {volume} {69}},\
  \bibinfo {pages} {094431} (\bibinfo {year} {2004})}\BibitemShut {NoStop}%
\bibitem [{\citenamefont {Schmidt}\ \emph {et~al.}(2003)\citenamefont
  {Schmidt}, \citenamefont {Monien},\ and\ \citenamefont
  {Uhrig}}]{Schmidt-M-U-03}%
  \BibitemOpen
  \bibfield  {author} {\bibinfo {author} {\bibfnamefont {K.~P.}\ \bibnamefont
  {Schmidt}}, \bibinfo {author} {\bibfnamefont {H.}~\bibnamefont {Monien}}, \
  and\ \bibinfo {author} {\bibfnamefont {G.~S.}\ \bibnamefont {Uhrig}},\ }\href
  {http://link.aps.org/abstract/PRB/v67/e184413} {\bibfield  {journal}
  {\bibinfo  {journal} {Phys. Rev. B}\ }\textbf {\bibinfo {volume} {67}},\
  \bibinfo {pages} {184413} (\bibinfo {year} {2003})}\BibitemShut {NoStop}%
\bibitem [{\citenamefont {Hakobyan}(2008)}]{Hakobyan-08}%
  \BibitemOpen
  \bibfield  {author} {\bibinfo {author} {\bibfnamefont {T.}~\bibnamefont
  {Hakobyan}},\ }\href {http://link.aps.org/doi/10.1103/PhysRevB.78.012407}
  {\bibfield  {journal} {\bibinfo  {journal} {Phys. Rev. B}\ }\textbf {\bibinfo
  {volume} {78}},\ \bibinfo {pages} {012407} (\bibinfo {year}
  {2008})}\BibitemShut {NoStop}%
\bibitem [{\citenamefont {Nishimoto}\ and\ \citenamefont
  {Arikawa}(2009)}]{Nishimoto-A-09}%
  \BibitemOpen
  \bibfield  {author} {\bibinfo {author} {\bibfnamefont {S.}~\bibnamefont
  {Nishimoto}}\ and\ \bibinfo {author} {\bibfnamefont {M.}~\bibnamefont
  {Arikawa}},\ }\href {http://link.aps.org/doi/10.1103/PhysRevB.79.113106}
  {\bibfield  {journal} {\bibinfo  {journal} {Phys. Rev. B}\ }\textbf {\bibinfo
  {volume} {79}},\ \bibinfo {pages} {113106} (\bibinfo {year}
  {2009})}\BibitemShut {NoStop}%
\bibitem [{\citenamefont {Hijii}\ \emph {et~al.}(2003)\citenamefont {Hijii},
  \citenamefont {Qin},\ and\ \citenamefont {Nomura}}]{Hijii-Q-N-03}%
  \BibitemOpen
  \bibfield  {author} {\bibinfo {author} {\bibfnamefont {K.}~\bibnamefont
  {Hijii}}, \bibinfo {author} {\bibfnamefont {S.}~\bibnamefont {Qin}}, \ and\
  \bibinfo {author} {\bibfnamefont {K.}~\bibnamefont {Nomura}},\ }\href
  {http://link.aps.org/doi/10.1103/PhysRevB.68.134403} {\bibfield  {journal}
  {\bibinfo  {journal} {Phys. Rev. B}\ }\textbf {\bibinfo {volume} {68}},\
  \bibinfo {pages} {134403} (\bibinfo {year} {2003})}\BibitemShut {NoStop}%
\bibitem [{\citenamefont {Lecheminant}\ and\ \citenamefont
  {Totsuka}(2005)}]{Lecheminant-T-05}%
  \BibitemOpen
  \bibfield  {author} {\bibinfo {author} {\bibfnamefont {P.}~\bibnamefont
  {Lecheminant}}\ and\ \bibinfo {author} {\bibfnamefont {K.}~\bibnamefont
  {Totsuka}},\ }\href {http://link.aps.org/abstract/PRB/v71/e020407} {\bibfield
   {journal} {\bibinfo  {journal} {Phys. Rev. B}\ }\textbf {\bibinfo {volume}
  {71}},\ \bibinfo {pages} {020407} (\bibinfo {year} {2005})}\BibitemShut
  {NoStop}%
\bibitem [{\citenamefont {Lecheminant}\ and\ \citenamefont
  {Totsuka}(2006)}]{Lecheminant-T-06-SU4}%
  \BibitemOpen
  \bibfield  {author} {\bibinfo {author} {\bibfnamefont {P.}~\bibnamefont
  {Lecheminant}}\ and\ \bibinfo {author} {\bibfnamefont {K.}~\bibnamefont
  {Totsuka}},\ }\href {http://link.aps.org/abstract/PRB/v74/e224426} {\bibfield
   {journal} {\bibinfo  {journal} {Phys. Rev. B}\ }\textbf {\bibinfo {volume}
  {74}},\ \bibinfo {pages} {224426} (\bibinfo {year} {2006})}\BibitemShut
  {NoStop}%
\bibitem [{\citenamefont {Song}\ \emph {et~al.}(2006)\citenamefont {Song},
  \citenamefont {Gu},\ and\ \citenamefont {Lin}}]{Song-G-L-06}%
  \BibitemOpen
  \bibfield  {author} {\bibinfo {author} {\bibfnamefont {J.-L.}\ \bibnamefont
  {Song}}, \bibinfo {author} {\bibfnamefont {S.-J.}\ \bibnamefont {Gu}}, \ and\
  \bibinfo {author} {\bibfnamefont {H.-Q.}\ \bibnamefont {Lin}},\ }\href
  {http://link.aps.org/abstract/PRB/v74/e155119} {\bibfield  {journal}
  {\bibinfo  {journal} {Phys. Rev. B}\ }\textbf {\bibinfo {volume} {74}},\
  \bibinfo {pages} {155119} (\bibinfo {year} {2006})}\BibitemShut {NoStop}%
\bibitem [{\citenamefont {Maruyama}\ \emph {et~al.}(2009)\citenamefont
  {Maruyama}, \citenamefont {Hirano},\ and\ \citenamefont
  {Hatsugai}}]{Maruyama-H-H-09}%
  \BibitemOpen
  \bibfield  {author} {\bibinfo {author} {\bibfnamefont {I.}~\bibnamefont
  {Maruyama}}, \bibinfo {author} {\bibfnamefont {T.}~\bibnamefont {Hirano}}, \
  and\ \bibinfo {author} {\bibfnamefont {Y.}~\bibnamefont {Hatsugai}},\ }\href
  {http://link.aps.org/doi/10.1103/PhysRevB.79.115107} {\bibfield  {journal}
  {\bibinfo  {journal} {Phys. Rev. B}\ }\textbf {\bibinfo {volume} {79}},\
  \bibinfo {pages} {115107} (\bibinfo {year} {2009})}\BibitemShut {NoStop}%
\bibitem [{\citenamefont {Arikawa}\ \emph {et~al.}(2009)\citenamefont
  {Arikawa}, \citenamefont {Tanaya}, \citenamefont {Maruyama},\ and\
  \citenamefont {Hatsugai}}]{Arikawa-T-M-H-09}%
  \BibitemOpen
  \bibfield  {author} {\bibinfo {author} {\bibfnamefont {M.}~\bibnamefont
  {Arikawa}}, \bibinfo {author} {\bibfnamefont {S.}~\bibnamefont {Tanaya}},
  \bibinfo {author} {\bibfnamefont {I.}~\bibnamefont {Maruyama}}, \ and\
  \bibinfo {author} {\bibfnamefont {Y.}~\bibnamefont {Hatsugai}},\ }\href
  {http://link.aps.org/doi/10.1103/PhysRevB.79.205107} {\bibfield  {journal}
  {\bibinfo  {journal} {Phys. Rev. B}\ }\textbf {\bibinfo {volume} {79}},\
  \bibinfo {pages} {205107} (\bibinfo {year} {2009})}\BibitemShut {NoStop}%
\bibitem [{\citenamefont {Li}\ \emph {et~al.}()\citenamefont {Li},
  \citenamefont {Shi}, \citenamefont {Liu}, ,\ and\ \citenamefont
  {Zhou}}]{Li-S-L-Z-11-unpub}%
  \BibitemOpen
  \bibfield  {author} {\bibinfo {author} {\bibfnamefont {S.-H.}\ \bibnamefont
  {Li}}, \bibinfo {author} {\bibfnamefont {Q.-Q.}\ \bibnamefont {Shi}},
  \bibinfo {author} {\bibfnamefont {J.-H.}\ \bibnamefont {Liu}}, , \ and\
  \bibinfo {author} {\bibfnamefont {H.-Q.}\ \bibnamefont {Zhou}},\ }\href
  {http://arxiv.org/abs/1105.3276} {\enquote {\bibinfo {title} {Ground-state
  fidelity and quantum criticality in a two-leg ladder with cyclic four-spin
  exchange},}\ }\bibinfo {note} {ArXiv 1105.3276}\BibitemShut {NoStop}%
\bibitem [{\citenamefont {Sachdev}\ and\ \citenamefont
  {Bhatt}(1990)}]{Sachdev-B-90}%
  \BibitemOpen
  \bibfield  {author} {\bibinfo {author} {\bibfnamefont {S.}~\bibnamefont
  {Sachdev}}\ and\ \bibinfo {author} {\bibfnamefont {R.~N.}\ \bibnamefont
  {Bhatt}},\ }\href {http://link.aps.org/abstract/PRB/v41/p9323} {\bibfield
  {journal} {\bibinfo  {journal} {Phys. Rev. B}\ }\textbf {\bibinfo {volume}
  {41}},\ \bibinfo {pages} {9323} (\bibinfo {year} {1990})}\BibitemShut
  {NoStop}%
\bibitem [{\citenamefont {Kolezhuk}(2007)}]{Kolezhuk-07}%
  \BibitemOpen
  \bibfield  {author} {\bibinfo {author} {\bibfnamefont {A.~K.}\ \bibnamefont
  {Kolezhuk}},\ }\href {http://link.aps.org/doi/10.1103/PhysRevLett.99.020405}
  {\bibfield  {journal} {\bibinfo  {journal} {Phys. Rev. Lett.}\ }\textbf
  {\bibinfo {volume} {99}},\ \bibinfo {pages} {020405} (\bibinfo {year}
  {2007})}\BibitemShut {NoStop}%
\bibitem [{\citenamefont {Sato}(2007)}]{Sato-07}%
  \BibitemOpen
  \bibfield  {author} {\bibinfo {author} {\bibfnamefont {M.}~\bibnamefont
  {Sato}},\ }\href {http://link.aps.org/abstract/PRB/v76/e054427} {\bibfield
  {journal} {\bibinfo  {journal} {Phys. Rev. B}\ }\textbf {\bibinfo {volume}
  {76}},\ \bibinfo {pages} {054427} (\bibinfo {year} {2007})}\BibitemShut
  {NoStop}%
\bibitem [{\citenamefont {Ho}(1998)}]{Ho-98}%
  \BibitemOpen
  \bibfield  {author} {\bibinfo {author} {\bibfnamefont {T.-L.}\ \bibnamefont
  {Ho}},\ }\href {http://link.aps.org/abstract/PRL/v81/p742} {\bibfield
  {journal} {\bibinfo  {journal} {Phys. Rev. Lett.}\ }\textbf {\bibinfo
  {volume} {81}},\ \bibinfo {pages} {742} (\bibinfo {year} {1998})}\BibitemShut
  {NoStop}%
\bibitem [{\citenamefont {Ohmi}\ and\ \citenamefont
  {Machida}(1998)}]{Ohmi-M-98}%
  \BibitemOpen
  \bibfield  {author} {\bibinfo {author} {\bibfnamefont {T.}~\bibnamefont
  {Ohmi}}\ and\ \bibinfo {author} {\bibfnamefont {K.}~\bibnamefont {Machida}},\
  }\href {http://jpsj.ipap.jp/link?JPSJ/67/1822/} {\bibfield  {journal}
  {\bibinfo  {journal} {J. Phys. Soc. Jpn.}\ }\textbf {\bibinfo {volume}
  {67}},\ \bibinfo {pages} {1822} (\bibinfo {year} {1998})}\BibitemShut
  {NoStop}%
\bibitem [{\citenamefont {Imambekov}\ \emph {et~al.}(2003)\citenamefont
  {Imambekov}, \citenamefont {Lukin},\ and\ \citenamefont
  {Demler}}]{Imambekov-L-D-03}%
  \BibitemOpen
  \bibfield  {author} {\bibinfo {author} {\bibfnamefont {A.}~\bibnamefont
  {Imambekov}}, \bibinfo {author} {\bibfnamefont {M.}~\bibnamefont {Lukin}}, \
  and\ \bibinfo {author} {\bibfnamefont {E.}~\bibnamefont {Demler}},\ }\href
  {http://link.aps.org/abstract/PRA/v68/e063602} {\bibfield  {journal}
  {\bibinfo  {journal} {Phys. Rev. {A}}\ }\textbf {\bibinfo {volume} {68}},\
  \bibinfo {pages} {063602} (\bibinfo {year} {2003})}\BibitemShut {NoStop}%
\bibitem [{\citenamefont {Kolezhuk}(1996)}]{Kolezhuk-96}%
  \BibitemOpen
  \bibfield  {author} {\bibinfo {author} {\bibfnamefont {A.~K.}\ \bibnamefont
  {Kolezhuk}},\ }\href {http://link.aps.org/abstract/PRB/v53/p318} {\bibfield
  {journal} {\bibinfo  {journal} {Phys. Rev. B}\ }\textbf {\bibinfo {volume}
  {53}},\ \bibinfo {pages} {318} (\bibinfo {year} {1996})}\BibitemShut
  {NoStop}%
\bibitem [{\citenamefont {Auerbach}(1994)}]{Auerbach-book}%
  \BibitemOpen
  \bibfield  {author} {\bibinfo {author} {\bibfnamefont {A.}~\bibnamefont
  {Auerbach}},\ }\href@noop {} {\emph {\bibinfo {title} {Interacting Electrons
  and Quantum Magnetism}}}\ (\bibinfo  {publisher} {Springer-Verlag},\ \bibinfo
  {year} {1994})\BibitemShut {NoStop}%
\bibitem [{\citenamefont {Wen}(2004)}]{Wen-book-04}%
  \BibitemOpen
  \bibfield  {author} {\bibinfo {author} {\bibfnamefont {X.~G.}\ \bibnamefont
  {Wen}},\ }\href@noop {} {\emph {\bibinfo {title} {Quantum Field Theory of
  Many-Body Systems}}}\ (\bibinfo  {publisher} {Oxford University Press},\
  \bibinfo {year} {2004})\BibitemShut {NoStop}%
\bibitem [{\citenamefont {Andreev}\ and\ \citenamefont
  {Grishchuk}(1985)}]{Andreev-G-85}%
  \BibitemOpen
  \bibfield  {author} {\bibinfo {author} {\bibfnamefont {A.}~\bibnamefont
  {Andreev}}\ and\ \bibinfo {author} {\bibfnamefont {I.}~\bibnamefont
  {Grishchuk}},\ }\href@noop {} {\bibfield  {journal} {\bibinfo  {journal}
  {Sov.Phys. JETP}\ }\textbf {\bibinfo {volume} {60}},\ \bibinfo {pages} {267}
  (\bibinfo {year} {1985})}\BibitemShut {NoStop}%
\bibitem [{\citenamefont {Totsuka}\ \emph {et~al.}()\citenamefont {Totsuka},
  \citenamefont {Capponi},\ and\ \citenamefont
  {Lecheminant}}]{Totsuka-C-L-unpub-12}%
  \BibitemOpen
  \bibfield  {author} {\bibinfo {author} {\bibfnamefont {K.}~\bibnamefont
  {Totsuka}}, \bibinfo {author} {\bibfnamefont {S.}~\bibnamefont {Capponi}}, \
  and\ \bibinfo {author} {\bibfnamefont {P.}~\bibnamefont {Lecheminant}},\
  }\href@noop {} {}\bibinfo {note} {In preparation}\BibitemShut {NoStop}%
\bibitem [{\citenamefont {Demler}\ and\ \citenamefont
  {Zhou}(2002)}]{Demler-Z-02}%
  \BibitemOpen
  \bibfield  {author} {\bibinfo {author} {\bibfnamefont {E.}~\bibnamefont
  {Demler}}\ and\ \bibinfo {author} {\bibfnamefont {F.}~\bibnamefont {Zhou}},\
  }\href {http://link.aps.org/abstract/PRL/v88/e163001} {\bibfield  {journal}
  {\bibinfo  {journal} {Phys. Rev. Lett.}\ }\textbf {\bibinfo {volume} {88}},\
  \bibinfo {pages} {163001} (\bibinfo {year} {2002})}\BibitemShut {NoStop}%
\bibitem [{\citenamefont {Zhou}(2003)}]{Zhou-review-03}%
  \BibitemOpen
  \bibfield  {author} {\bibinfo {author} {\bibfnamefont {F.}~\bibnamefont
  {Zhou}},\ }\href@noop {} {\bibfield  {journal} {\bibinfo  {journal} {Int. J.
  Mod. Phys. B}\ }\textbf {\bibinfo {volume} {17}},\ \bibinfo {pages} {2643}
  (\bibinfo {year} {2003})}\BibitemShut {NoStop}%
\bibitem [{\citenamefont {Essler}\ \emph {et~al.}(2009)\citenamefont {Essler},
  \citenamefont {Shlyapnikov},\ and\ \citenamefont {Tsvelik}}]{Essler-S-T-09}%
  \BibitemOpen
  \bibfield  {author} {\bibinfo {author} {\bibfnamefont {F.~H.~L.}\
  \bibnamefont {Essler}}, \bibinfo {author} {\bibfnamefont {G.~V.}\
  \bibnamefont {Shlyapnikov}}, \ and\ \bibinfo {author} {\bibfnamefont {A.~M.}\
  \bibnamefont {Tsvelik}},\ }\href
  {http://stacks.iop.org/1742-5468/2009/P02027} {\bibfield  {journal} {\bibinfo
   {journal} {J. Stat. Mech.: Theory and Experiment}\ }\textbf {\bibinfo
  {volume} {2009}},\ \bibinfo {pages} {P02027} (\bibinfo {year}
  {2009})}\BibitemShut {NoStop}%
\bibitem [{\citenamefont {Zamolodchikov}\ and\ \citenamefont
  {Zamolodchikov}(1979)}]{Zamolodchikov-Z-79}%
  \BibitemOpen
  \bibfield  {author} {\bibinfo {author} {\bibfnamefont {A.~B.}\ \bibnamefont
  {Zamolodchikov}}\ and\ \bibinfo {author} {\bibfnamefont {A.~B.}\ \bibnamefont
  {Zamolodchikov}},\ }\href {\doibase 10.1016/0003-4916(79)90391-9} {\bibfield
  {journal} {\bibinfo  {journal} {Annals of Physics}\ }\textbf {\bibinfo
  {volume} {120}},\ \bibinfo {pages} {253 } (\bibinfo {year}
  {1979})}\BibitemShut {NoStop}%
\bibitem [{\citenamefont {S\'{e}n\'{e}chal}(1995)}]{Senechal-95}%
  \BibitemOpen
  \bibfield  {author} {\bibinfo {author} {\bibfnamefont {D.}~\bibnamefont
  {S\'{e}n\'{e}chal}},\ }\href
  {http://link.aps.org/doi/10.1103/PhysRevB.52.15319} {\bibfield  {journal}
  {\bibinfo  {journal} {Phys. Rev. B}\ }\textbf {\bibinfo {volume} {52}},\
  \bibinfo {pages} {15319} (\bibinfo {year} {1995})}\BibitemShut {NoStop}%
\bibitem [{\citenamefont {Sierra}(1996)}]{Sierra-96}%
  \BibitemOpen
  \bibfield  {author} {\bibinfo {author} {\bibfnamefont {G.}~\bibnamefont
  {Sierra}},\ }\href {http://stacks.iop.org/0305-4470/29/3299} {\bibfield
  {journal} {\bibinfo  {journal} {J. Phys. A: Mathematical and General}\
  }\textbf {\bibinfo {volume} {29}},\ \bibinfo {pages} {3299} (\bibinfo {year}
  {1996})}\BibitemShut {NoStop}%
\bibitem [{Note1()}]{Note1}%
  \BibitemOpen
  \bibinfo {note} {The explicit form of the normal modes around $k=\pi $ shows
  that the corresponding fluctuation is of the form $\protect \mathbf {A}\DOTSB
  \mapstochar \rightarrow \protect \text {e}^{i\delta \phi }\protect \mathbf
  {A}$.}\BibitemShut {Stop}%
\bibitem [{\citenamefont {Hikihara}\ and\ \citenamefont
  {Yamamoto}(2008)}]{Hikihara-Y-08}%
  \BibitemOpen
  \bibfield  {author} {\bibinfo {author} {\bibfnamefont {T.}~\bibnamefont
  {Hikihara}}\ and\ \bibinfo {author} {\bibfnamefont {S.}~\bibnamefont
  {Yamamoto}},\ }\href {\doibase 10.1143/JPSJ.77.014709} {\bibfield  {journal}
  {\bibinfo  {journal} {J. Phys. Soc. Jpn.}\ }\textbf {\bibinfo {volume}
  {77}},\ \bibinfo {pages} {014709} (\bibinfo {year} {2008})}\BibitemShut
  {NoStop}%
\bibitem [{Note2()}]{Note2}%
  \BibitemOpen
  \bibinfo {note} {In 1D, the projection of $\protect \mathbf {B}$ onto the
  plane perpendicular to $H$ cannot order and only the $z$-component takes a
  finite expectation value; in higher dimension, on the other hand, it is
  possible that both $\protect \mathbf {A}$ and $\protect \mathbf {B}$ exhibit
  long-range order.}\BibitemShut {Stop}%
\bibitem [{\citenamefont {Sakai}\ and\ \citenamefont
  {Hasegawa}(1999)}]{Sakai-H-99}%
  \BibitemOpen
  \bibfield  {author} {\bibinfo {author} {\bibfnamefont {T.}~\bibnamefont
  {Sakai}}\ and\ \bibinfo {author} {\bibfnamefont {Y.}~\bibnamefont
  {Hasegawa}},\ }\href {http://link.aps.org/doi/10.1103/PhysRevB.60.48}
  {\bibfield  {journal} {\bibinfo  {journal} {Phys. Rev. B}\ }\textbf {\bibinfo
  {volume} {60}},\ \bibinfo {pages} {48} (\bibinfo {year} {1999})}\BibitemShut
  {NoStop}%
\bibitem [{\citenamefont {Nakasu}\ \emph {et~al.}(2001)\citenamefont {Nakasu},
  \citenamefont {Totsuka}, \citenamefont {Hasegawa}, \citenamefont {Okamoto},\
  and\ \citenamefont {Sakai}}]{Nakasu-T-H-O-S-01}%
  \BibitemOpen
  \bibfield  {author} {\bibinfo {author} {\bibfnamefont {A.}~\bibnamefont
  {Nakasu}}, \bibinfo {author} {\bibfnamefont {K.}~\bibnamefont {Totsuka}},
  \bibinfo {author} {\bibfnamefont {Y.}~\bibnamefont {Hasegawa}}, \bibinfo
  {author} {\bibfnamefont {K.}~\bibnamefont {Okamoto}}, \ and\ \bibinfo
  {author} {\bibfnamefont {T.}~\bibnamefont {Sakai}},\ }\href
  {http://stacks.iop.org/0953-8984/13/i=33/a=321} {\bibfield  {journal}
  {\bibinfo  {journal} {J. Phys.: Condensed Matter}\ }\textbf {\bibinfo
  {volume} {13}},\ \bibinfo {pages} {7421} (\bibinfo {year}
  {2001})}\BibitemShut {NoStop}%
\bibitem [{\citenamefont {Romh{\'{a}}nyi}\ \emph {et~al.}(2011)\citenamefont
  {Romh{\'{a}}nyi}, \citenamefont {Totsuka},\ and\ \citenamefont
  {Penc}}]{Romhanyi-T-P-11}%
  \BibitemOpen
  \bibfield  {author} {\bibinfo {author} {\bibfnamefont {J.}~\bibnamefont
  {Romh{\'{a}}nyi}}, \bibinfo {author} {\bibfnamefont {K.}~\bibnamefont
  {Totsuka}}, \ and\ \bibinfo {author} {\bibfnamefont {K.}~\bibnamefont
  {Penc}},\ }\href {http://link.aps.org/doi/10.1103/PhysRevB.83.024413}
  {\bibfield  {journal} {\bibinfo  {journal} {Phys. Rev. B}\ }\textbf {\bibinfo
  {volume} {83}},\ \bibinfo {pages} {024413} (\bibinfo {year}
  {2011})}\BibitemShut {NoStop}%
\bibitem [{\citenamefont {White}(1992)}]{White-92}%
  \BibitemOpen
  \bibfield  {author} {\bibinfo {author} {\bibfnamefont {S.~R.}\ \bibnamefont
  {White}},\ }\href {http://link.aps.org/doi/10.1103/PhysRevLett.69.2863}
  {\bibfield  {journal} {\bibinfo  {journal} {Phys. Rev. Lett.}\ }\textbf
  {\bibinfo {volume} {69}},\ \bibinfo {pages} {2863} (\bibinfo {year}
  {1992})}\BibitemShut {NoStop}%
\bibitem [{\citenamefont {Haldane}(1981)}]{Haldane-PRL-81}%
  \BibitemOpen
  \bibfield  {author} {\bibinfo {author} {\bibfnamefont {F.~D.~M.}\
  \bibnamefont {Haldane}},\ }\href
  {http://link.aps.org/doi/10.1103/PhysRevLett.47.1840} {\bibfield  {journal}
  {\bibinfo  {journal} {Phys. Rev. Lett.}\ }\textbf {\bibinfo {volume} {47}},\
  \bibinfo {pages} {1840} (\bibinfo {year} {1981})}\BibitemShut {NoStop}%
\bibitem [{Note3()}]{Note3}%
  \BibitemOpen
  \bibinfo {note} {After neglecting the longitudinal fluctuations, the charge
  part ${\protect \cal H}_{\protect \text {charge}}$ (\ref
  {eqn:classical-AB-charge}) merely contributes a constant.}\BibitemShut
  {Stop}%
\end{thebibliography}%
\end{document}